\documentclass[%
 prfluids,
 superscriptaddress,
 amsmath,amssymb,
 aps,
onecolumn,
preprint,  %% for preprint with double space
titlepage, %% for preprint with double space
longbibliography
]{revtex4-2}

\usepackage{graphicx}
\usepackage{dcolumn}% Align table columns on decimal point
\usepackage{bm}% bold math
\usepackage{natbib}
\usepackage{amsfonts}
\usepackage{mathrsfs}
\usepackage{color}
\usepackage{comment}

\usepackage{CJK}

\newcommand{\Tay}{\mbox{$\mathrm{Ta}$}}
\renewcommand{\d}{\mathrm{d}}
\newcommand{\D}{\mathrm{D}}
\newcommand{\Rey}{\mbox{$\mathrm{Re}$}}
\newcommand{\Sto}{\mbox{$\mathrm{St}$}}

\begin{document}

%\begin{CJK*}{UTF8}{min}

%\preprint{APS/123-QED}

\title{
Explanation of constant mean angular momentum\\
in high-Reynolds-number Taylor--Couette turbulence\\
in terms of history effects
}% Force line breaks with \\

\author{Kazuhiro Inagaki}
 \email{Current address: Department of Physics \& Research and Education Center for Natural Sciences, Hiyoshi Campus, Keio University, Yokohama 223-8521, Kanagawa, Japan
 inagaki@keio.jp}
\affiliation{%
Department of Mechanical Engineering, Doshisha University, Kyotanabe 610-0394, Kyoto, Japan
}

\author{Yasufumi Horimoto}
 \email{horimoto@mech.kindai.ac.jp}
\affiliation{%
Laboratory of Complex Flows, Faculty of Science and Engineering, Kindai University, Higashiosaka 577-8502, Osaka, Japan
}%

\date{\today}% It is always \today, today,

\allowdisplaybreaks[1]

\begin{abstract}

This study discusses the mechanism of the emergence of nearly constant mean angular momentum profiles, which are widely observed in curved turbulent flows including the bulk region of Taylor--Couette (TC) flows.
For high-Reynolds-number TC flows where the inner and outer cylinders are weakly counter-rotating and co-rotating, both the bulk and boundary layers become turbulent without Taylor rolls, referred to as the featureless ultimate regime (UR).
Thus, we utilize the Reynolds-averaged Navier--Stokes (RANS) equations to explain the mechanism of the nearly constant mean angular momentum.
High-Reynolds-number experiments of TC turbulence are performed for reference, where the angular velocity ratio $a = -\omega_\mathrm{out}/\omega_\mathrm{in}$ is in the range $-0.5 \le a \le 0.1$.
Verification of the RANS based on the conventional algebraic Reynolds stress model suggests that convection of the Reynolds stress is essential for predicting the angular momentum profile.
This indicates that the physical origin of the nearly constant angular momentum is the history effect of the Reynolds stress.
To rigorously incorporate the convection effect into the Reynolds stress, we employ the Jaumann derivative as a covariant time derivative.
The model that takes into account the history effect involving the normal stress difference successfully predicts the nearly constant mean angular momentum in the co-rotating cases.
This study suggests the significance of the history effects for understanding curved or rotating turbulent flows in terms of the statistical analysis.

\end{abstract}

\pacs{Valid PACS appear here}% PACS, the Physics and Astronomy
\maketitle

%\end{CJK*}

\section{Introduction}\label{sec:intro}

The Taylor--Couette (TC) flow, driven by the friction of two independently rotating concentric cylinders, has long been studied to investigate the effects of rotation or flow curvature on turbulence.
Over the last two decades, exploration of high-Reynolds-number TC turbulence has progressed both experimentally and numerically \citep{grossmannetal2016}. 
At sufficiently high Reynolds numbers, both the bulk and boundary layers on the cylinders become turbulent \citep{hk2011,huismanetal2013,ostillamonicoetal2014a,ostillamonicoetal2014b,ostillamonicoetal2014c,chouippeetal2014,berghoutetal2020}. 
This high-Reynolds-number regime with turbulent boundary layers is referred to as the ultimate regime (UR).
Furthermore, for cases in which two cylinders are weakly counter-rotating and co-rotating, the so-called Taylor rolls, which indicate the gap-width-size mean azimuthal vortices that remain after time averaging, vanish; thus, the flows lead to a featureless UR \citep{ostillamonicoetal2014c}. 
Featureless turbulence with co-rotating inner and outer boundaries is often studied in the context of angular momentum transport in astrophysical objects such as accretion disks \citep{dubrulleetal2005,schartmanetal2009,jb2013}.
For the featureless UR, the statistics of TC flow yield a one-dimensional function of radial position.
Such one-dimensional statistics of turbulent flows may be predicted using a relatively simple Reynolds-averaged Navier--Stokes (RANS) model. 
It is valuable to discuss the physical origin of the statistics of curved or rotating turbulent flows in terms of the RANS models because they provide a physical interpretation of high-Reynolds-number turbulence phenomena, such as meteorological or astrophysical flows.
The RANS models have several deficiencies in predicting complex turbulence, including rotating flows \citep{popebook}.
Several theoretical studies showed that some of these deficiencies can be resolved by modifying the model of the Reynolds stress \citep{yoshizawabook}.
For example, \citet{speziale1987} demonstrated that the nonlinear eddy-viscosity models successfully predict the secondary flow in a square duct.
The success originates from the prediction of the normal stress anisotropy, which is ignored in simple eddy-viscosity models.
In addition, RANS models have the potential to predict a non-trivial phenomenon: e.g., mean flow generation through turbulent kinetic helicity in rotating turbulence \citep{yb2016,inagakietal2017}.
These studies demonstrated that the RANS model accompanied by the term involving the spatially inhomogeneous distribution of turbulent kinetic helicity coupled with large-scale rotation, which was first derived theoretically by \citet{yy1993}, accounts for the mean flow generation phenomenon.
Thus, resolving the deficiencies of RANS models by taking into account a new term leads to the underlying physics in turbulent flows.

There exist several theoretically rigorous results for torque scaling or bounds for the Reynolds stress based on the conservation equations \citep{eckhardtetal2007,busse1996}.
Although these studies provided significant information about the angular velocity or momentum flux due to turbulence, they did not explain the velocity profiles.
This situation is similar to the parallel shear flows; to derive the logarithmic velocity profile, we require several additional assumptions beyond the Reynolds stress scaling by the friction velocity.
\citet{busse1972} provided an asymptotic velocity profile for the TC flows using the upper bound theory.
However, the profile is not accurate for the cases in which both the inner and outer cylinders rotate \citep{vangilsetal2012}.
The mean azimuthal velocity profile is closely related to the optimal angular velocity transport \citep{grossmannetal2016}.
Thus, constructing models that predict the velocity profiles is beneficial.

Curved or rotating turbulent flows exhibit characteristic mean velocity profiles. For turbulent TC flows, a nearly constant mean angular momentum appears in the bulk region, as shown in previous experiments \citep{st1982,ls1999,froitzheimetal2017,ezetaetal2018}, numerical simulations \citep{dong2007,be2013,brauckmannetal2016,froitzheimetal2019,chengetal2020}, and theoretical analysis \citep{deguchi2023}. 
\citet{brauckmannetal2016} showed that the nearly constant mean angular momentum is robust to a wide range of angular velocity ratios of the outer cylinder to the inner cylinder and is observed for weakly counter-rotating and co-rotating cases. 
A nearly constant mean angular momentum was also observed in curved channel flows with strong curvature \citep{wattendorf1935,nk2004,brethouwer2022}. 
A similar characteristic statistical property of rotating turbulent flows is a nearly zero mean absolute vorticity \citep{johnstonetal1972,tanakaetal2000,hamba2006,grundestametal2008,xiaetal2016,ka2016jfm}. 
Both the constant mean angular momentum and zero mean absolute vorticity conform to neutral stability \citep{cambonetal1994,brauckmannetal2016,brethouwer2022}. 
Predicting the nearly zero mean absolute vorticity has been a good benchmark for RANS modeling because it cannot be reproduced using standard eddy-viscosity models~\citep{gs1993,wj2000,wj2002,gw2004,hamba2006}. 
In other words, considering a model expression that explains characteristic statistics, such as the nearly zero mean absolute vorticity, will provide a physical essence for representing them.
To express the effects of the frame rotation or flow curvature, the convection term of the Reynolds stress has often been considered for RANS modeling \cite{girimaji1997,wj2002,gw2004}.
\citet{hamba2006,hamba2017} discussed the effects of the convection of the Reynolds stress as a time history effect via the Green's function.
Thus, the history effect of the Reynolds stress is a candidate for expressing the flow curvature effects, leading to the explanation of the nearly constant mean angular momentum states.
This study demonstrates the significance of the history effect of the Reynolds stress involving the normal stress difference for the physical understanding of the states.
Furthermore, we formulate the history effects independently of the coordinate, by using the Jaumann derivative as a covariant time derivative.

The remainder of this paper is organized as follows. 
In Sec.~\ref{sec:equations}, we summarize the basic equations for the mean velocity and Reynolds stress. 
We also derive an algebraic model expression of Reynolds stress involving the convection effects in cylindrical coordinates. 
The performance of the derived model is discussed and compared with experimental results for the featureless UR of the TC turbulence in Sec.~\ref{sec:verification}. 
We discuss the physical essence of the model for predicting the nearly constant mean angular momentum in Sec.~\ref{sec:discussion}. 
Specifically, we provide a model expression for the Reynolds stress incorporating the history effects using the Jaumann derivative.
Finally, the performance and physical understanding of the resultant model are discussed.
The conclusions are presented in Sec.~\ref{sec:conclusions}.

\section{Basic equations and covariance}\label{sec:equations}

\subsection{Governing equations}\label{sec:governingequations}

We usually employ cylindrical coordinates for TC flows, whereas we employ Cartesian coordinates for spanwise rotating turbulent channel flows in which the nearly zero mean absolute vorticity is observed.
To systematically discuss the physics in different coordinates, a form-invariant notation of equations may be preferable (see e.g., Refs.~\citep{frewer2009,ariki2015}).
Therefore, we write the governing equations using the covariant derivative.

The continuity and the $i$th component of the Navier--Stokes equations for incompressible flows in an inertial frame are as follows \citep{frewer2009,ariki2015,kajishimatairabook}:
\begin{gather}
    \nabla_i u_i = 0, 
    \label{eq:continuityeq} \\
    \frac{\partial u_i}{\partial t} = - u_j \nabla_j u_i - \nabla_i p  + \nu \nabla_j \nabla_j u_i,
    \label{eq:nseq}
\end{gather}
where $u_i$, $p$, and $\nu$ denote the velocity field, pressure divided by density, kinematic viscosity, and metric tensor, respectively.
For cylindrical coordinates, the subscript $i$ or $j$ is $r$, $\theta$, or $z$ where $r$, $\theta$, and $z$ denote the radial, azimuthal, and axial directions, respectively.
Here, we used the covariant derivative $\nabla_i$ because the governing equations in cylindrical coordinates can be expressed in the same form as those in Cartesian coordinates.
In cylindrical coordinates, we have (see Appendix~\ref{sec:covariantderivative} for the derivation)
\begin{align}
    \begin{bmatrix}
    \nabla_r u_r & \nabla_\theta u_r & \nabla_z u_r \\
    \nabla_r u_\theta & \nabla_\theta u_\theta & \nabla_z u_\theta \\
    \nabla_r u_z & \nabla_\theta u_z & \nabla_z u_z
    \end{bmatrix}
    = 
    \begin{bmatrix}
    \displaystyle \frac{\partial u_r}{\partial r} & 
    \displaystyle \frac{1}{r} \frac{\partial u_r}{\partial \theta} - \frac{u_\theta}{r} &
    \displaystyle \frac{\partial u_r}{\partial z} \\
    \displaystyle \frac{\partial u_\theta}{\partial r} &
    \displaystyle \frac{1}{r} \frac{\partial u_\theta}{\partial \theta} + \frac{u_r}{r} &
    \displaystyle \frac{\partial u_\theta}{\partial z} \\
    \displaystyle \frac{\partial u_z}{\partial r} &
    \displaystyle \frac{1}{r} \frac{\partial u_z}{\partial \theta} &
    \displaystyle \frac{\partial u_z}{\partial z} 
    \end{bmatrix}.
    \label{eq:covariantvelocitygradient}
\end{align}
As a result, the continuity equation and each component of the nonlinear term of the Navier--Stokes equations in cylindrical coordinates in the inertial frame can be written as follows:
\begin{subequations}
\begin{align}
    \nabla_i u_i &
    = \frac{\partial u_r}{\partial r} 
    + \frac{1}{r} \frac{\partial u_\theta}{\partial \theta} + \frac{u_r}{r}
    + \frac{\partial u_z}{\partial z}
    = \frac{1}{r} \frac{\partial}{\partial r} (ru_r)
    + \frac{1}{r} \frac{\partial u_\theta}{\partial \theta}
    + \frac{\partial u_z}{\partial z} = 0, 
    \label{eq:continuityequation} \\
    u_j \nabla_j u_r & 
    = u_r \frac{\partial u_r}{\partial r} 
    + u_\theta \left( \frac{1}{r} \frac{\partial u_r}{\partial \theta} - \frac{u_\theta}{r} \right)
    + u_z \frac{\partial u_r}{\partial z} 
    \nonumber \\
    & = \frac{1}{r} \frac{\partial}{\partial r} (ru_r^2) 
    + \frac{1}{r} \frac{\partial}{\partial \theta} (u_\theta u_r)
    + \frac{\partial}{\partial z} (u_z u_r)
    - \frac{(u_\theta^2)}{r},
    \label{eq:rnonlinearterm} \\
    u_j \nabla_j u_\theta &
    = u_r \frac{\partial u_\theta}{\partial r} 
    + u_\theta \left( \frac{1}{r} \frac{\partial u_\theta}{\partial \theta} + \frac{u_r}{r} \right)
    + u_z \frac{\partial u_\theta}{\partial z}
    \nonumber \\
    & = \frac{1}{r} \frac{\partial}{\partial r} (ru_r u_\theta) 
    + \frac{1}{r} \frac{\partial}{\partial \theta} (u_\theta^2)
    + \frac{\partial}{\partial z} (u_z u_\theta)
    + \frac{u_r u_\theta}{r},
    \label{eq:thetanonlinearterm} \\
    u_j \nabla_j u_z &
    = u_r \frac{\partial u_z}{\partial r} 
    + u_\theta \frac{1}{r} \frac{\partial u_z}{\partial \theta}
    + u_z \frac{\partial u_z}{\partial z}
    \nonumber \\
    & = \frac{1}{r} \frac{\partial}{\partial r} (ru_r u_z) 
    + \frac{1}{r} \frac{\partial}{\partial \theta} (u_\theta u_z)
    + \frac{\partial}{\partial z} (u_z^2),
    \label{eq:znonlinearterm}
\end{align}
\end{subequations}
The fourth terms on the right-hand sides of Eqs.~(\ref{eq:rnonlinearterm}) and (\ref{eq:thetanonlinearterm}) originate from the curvature of the coordinate system. 
Hence, part of the flow curvature effect on the Reynolds stress results from the convection term.

\subsection{Transformation rules}\label{sec:transformationrules}

To discuss the relationship between the nearly zero mean absolute vorticity and constant angular momentum, let us summarize the transformation rules for velocity fields.

\subsubsection{Velocity field and Reynolds stress}\label{sec:transformationofvelocityandreynoldsstress}

Let us consider the transformation between two coordinate systems, $( \tilde{t}, \tilde{\bm{x}} )$ and $( t, \bm{x} )$, defined as
\begin{align}
    \tilde{x}_a (t,\bm{x})
    = Q_{ai} (t, \bm{x}) [x_i + x_i^\mathrm{O}(t)]
    = \frac{\partial \tilde{x}_a}{\partial x_i} [x_i + x_i^\mathrm{O}(t)],
\end{align}
where $x_j^\mathrm{O}(t)$ denotes the spatial shift of the origin, which depends only on time.
We only consider the non-relativistic case where $\tilde{t}=t$.
The transformation of velocity fields yields \citep{frewer2009,ariki2015}
\begin{align}
    \widetilde{u}_a (\tilde{t},\tilde{\bm{x}}) 
    = \frac{\partial \tilde{x}_a}{\partial x_i} u_i (t,\bm{x}) 
    + \frac{\partial \tilde{x}_a}{\partial t}
    = \frac{\partial \tilde{x}_a}{\partial x_i} \left[ u_i (t,\bm{x}) 
    + \frac{\d x_i^\mathrm{O}}{\d t} \right]
    - \widetilde{\Omega}^\mathrm{F}_{ai} \tilde{x}_i,
    \label{eq:velocitytransformation}
\end{align}
where
\begin{align}
    \widetilde{\Omega}^\mathrm{F}_{ij} 
    = \frac{\partial \tilde{x}_i}{\partial x_\ell} \frac{\partial^2 x_\ell}{\partial t \partial \tilde{x}_j}
    = Q_{i\ell} \frac{\partial Q^{-1}_{\ell j}}{\partial t}.
    \label{eq:rotationrate}
\end{align}
Note that $Q^{-1}_{i j}$ is the inverse of $Q_{i j}$ that satisfies $Q^{-1}_{i \ell} Q_{\ell j} = Q_{i \ell} Q^{-1}_{\ell j} = \delta_{ij}$ where $\delta_{ij}$ is Kronecker's delta.
In Eq.~(\ref{eq:velocitytransformation}), $\d x_j^\mathrm{O}/\d t$ denotes the Galilean boost or the relative velocity of the $( \tilde{t}, \tilde{\bm{x}} )$ system to the $( t, \bm{x} )$ system and $\widetilde{\Omega}^\mathrm{F}_{i j}$ denotes the rotation rate of the $( \tilde{t}, \tilde{\bm{x}} )$ system relative to $( t, \bm{x} )$ observed in the $( \tilde{t}, \tilde{\bm{x}} )$ system.

To derive the transformation rules for RANS, we consider the Reynolds or ensemble average: $f= F + f'$ where $f$ denotes basic flow variables such as $u_i$ or $p$, $F = \langle f \rangle$, and $\langle \cdot \rangle$ denotes the ensemble average. 
Considering the coordinate transformation of the velocity field, the Galilean boost $\d x^\mathrm{O}_i/\d t$ and the frame rotation $\tilde{\Omega}^\mathrm{F}_{ij}$ are included in that of the mean velocity.
Thus, the transformation rule for the velocity fluctuation $u_i'$ yields
\begin{align}
    \tilde{u}_a' (\tilde{t},\tilde{\bm{x}}) 
    = \frac{\partial \tilde{x}_a}{\partial x_i} u_i' (t,\bm{x}).
\end{align}
Therefore, the transformation of the Reynolds stress $R_{ij} = \langle u_i' u_j' \rangle$ reads
\begin{align}
    \widetilde{R}_{ab} = \langle \tilde{u}_a' \tilde{u}_b' \rangle
    = \frac{\partial \tilde{x}_a}{\partial x_i} \frac{\partial \tilde{x}_b}{\partial x_j} 
    \langle u_i' u_j' \rangle
    = \frac{\partial \tilde{x}_a}{\partial x_i} \frac{\partial \tilde{x}_b}{\partial x_j}
    R_{ij}.
    \label{eq:covarianceofreynoldsstress}
\end{align}
The quantities that obey the transformation rule expressed by Eq.~(\ref{eq:covarianceofreynoldsstress}) are called tensors, and this transformation property is referred to as covariance or form invariance (see Refs.~\citep{speziale1979,frewer2009,ariki2015} for the details of the coordinate transformation in fluid mechanics).
Note that any model expressions for the Reynolds stress must satisfy the transformation rule given by Eq.~(\ref{eq:covarianceofreynoldsstress}).
In turbulence modeling, we express the Reynolds stress in terms of statistical quantities such as the turbulent kinetic energy, mean strain rate, and so on.
Turbulence models often provide a physical interpretation of phenomena, as the eddy viscosity models explain the enhancement of mixing due to velocity fluctuation.
If a model violates the transformation rule expressed by Eq.~(\ref{eq:covarianceofreynoldsstress}), the interpretation of phenomena based on that model depends on the coordinate system.
It is valuable to discuss the covariant turbulence models for understanding the physics in a coordinate-independent manner.

\subsubsection{Time derivative of tensors}\label{sec:transformationoftimederivatives}

We can derive an algebraic model expression for the Reynolds stress based on its transport equations (see e.g., Ref.~\citep{popebook}).
Specifically, we often consider the time derivative of the anisotropy tensor $b_{ij}$ defined by
\begin{align}
    b_{ij} = \frac{R_{ij}}{K} - \frac{2}{3} \delta_{ij},
\end{align}
with the turbulent kinetic energy $K (= R_{ii}/2)$.
Of course, $b_{ij}$ forms a tensor as the Reynolds stress expressed by Eq.~(\ref{eq:covarianceofreynoldsstress}) does.
However, the Lagrangian or material derivative of tensors does not form a tensor; namely, the transformation of the Lagrangian derivative of $b_{ij}$ along the mean velocity yields
\begin{align}
    \frac{\D \tilde{b}_{ab}}{\D \tilde{t}}
    = \left( \frac{\partial}{\partial \tilde{t}} + \widetilde{U}_c \widetilde{\nabla}_c \right) \tilde{b}_{ab}
    \neq \frac{\partial \tilde{x}_a}{\partial x_i} \frac{\partial \tilde{x}_b}{\partial x_j} 
    \left( \frac{\partial}{\partial t} + U_\ell \nabla_\ell \right) b_{ij},
    \label{eq:lagrangianderivativenottensor}
\end{align}
where $\D/\D t (=\partial/\partial t + U_i \nabla_i)$ denotes the Lagrangian derivative along the mean velocity.
Note that we define the Lagrangian derivative using the covariant derivative $\nabla_i$.
The inconsistency in the covariance may provide a coordinate-dependent understanding of phenomena.
Thus, interpreting the history effects on tensors based on the Lagrangian derivative can decrease physical understanding.

\subsection{Constant mean angular momentum and zero mean absolute vorticity}\label{sec:constantangularmomentum}

In turbulent TC flows, the mean velocity profile in the bulk region is often approximated by a nearly constant mean angular momentum $rU_\theta \simeq \text{const.}$ 
\citep{st1982,ls1999,froitzheimetal2017,ezetaetal2018,dong2007,be2013,brauckmannetal2016,froitzheimetal2019,chengetal2020,deguchi2023}. 
A nearly constant mean angular momentum has also been observed in curved turbulent channel flows with strong curvature \citep{wattendorf1935,nk2004,brethouwer2022}. 
\citet{brauckmannetal2016} and \citet{brethouwer2022} discussed the relationship between the constant mean angular momentum and zero mean absolute vorticity in rotating turbulent shear flows because both correspond to neutral stability. 
A nearly zero mean absolute vorticity state has been widely observed in the bulk region of rotating turbulent channel or plane Couette flows \citep{johnstonetal1972,tanakaetal2000,hamba2006,grundestametal2008,xiaetal2016,ka2016jfm,ka2016}. 

Here, we show that the constant mean angular momentum in circular flows corresponds exactly to zero mean absolute vorticity. 
Absolute vorticity is the covariant form of vorticity in a rotating frame.
For example, considering the coordinate transformation from an inertial frame $(t,\bm{x})$ to a rotating frame $(t^\dag,\bm{x}^\dag)$ with a constant angular velocity $\Omega^{\mathrm{F}\dag}_i = -\epsilon_{ij\ell} \Omega^{\mathrm{F}\dag}_{j\ell}/2$, where $\epsilon_{ij\ell}$ is Levi--Civita symbol, the vorticity tensor obeys the following transformation rule \citep{wh2003,gw2004,hamba2006jfm,ariki2015}:
\begin{align}
    w^{\mathrm{A}\dag}_{ab} = w^\dag_{ab} + \Omega^{\mathrm{F}\dag}_{ab}
    = \frac{\partial x^\dag_a}{\partial x_i} \frac{\partial x^\dag_b}{\partial x_j} w_{ij},
\end{align}
where $w^{\mathrm{A}\dag}_{ij}$ denotes the absolute vorticity tensor. 
In addition, $w_{ij}$ and $w^\dag_{ij}$ represent the vorticity matrices in the inertial and rotating frames, respectively, and are defined as
\begin{align}
    w_{ij} = \frac{1}{2} ( \nabla_j u_i - \nabla_i u_j), \ \ 
    w^\dag_{ij} = \frac{1}{2} ( \nabla^\dag_j u^\dag_i - \nabla^\dag_i u^\dag_j).
    \label{eq:definitionofvorticity}
\end{align}
Therefore, the zero mean absolute vorticity in the rotating frame $W^{\mathrm{A}\dag}_{ij} = \langle w^{\mathrm{A}\dag}_{ij} \rangle = 0$ coincides with the zero mean vorticity in the inertial frame $W_{ij} = \langle w_{ij} \rangle = 0$. For a cylindrical coordinate with $(U_r,U_\theta,U_z) = (0, U_\theta(r),0)$, the zero mean vorticity yields
\begin{align}
   W_{r\theta} = \frac{1}{2} \left( -\frac{U_\theta}{r} -\frac{\d U_\theta}{\d r} \right)
    = - \frac{1}{2r} \frac{\d}{\d r} (rU_\theta) = 0 
    \ \iff \
    rU_\theta = \mathscr{L} = \text{const.},
    \label{eq:constantangularmomentum}
\end{align}
which represents the constant mean angular momentum.

It is worth noting that the standard linear eddy-viscosity models cannot predict the nearly zero mean absolute vorticity because they do not involve the effects of frame rotation.
To predict the nearly zero mean absolute vorticity in the RANS model, we must properly take into account the effects of frame rotation \citep{wj2000,wj2002,gw2004,hamba2006}.
In other words, the proper way to account for the frame rotation effects provides knowledge about the physical origin of the nearly zero mean absolute vorticity.
Therefore, we expect that we can explain the physics of the nearly constant mean angular momentum by considering proper RANS models.

\subsection{Basic equations for TC flow in featureless UR}\label{sec:basicequationsfortcflow}

For the TC flow, the ensemble average equals the average over the azimuthal direction and time. 
For the featureless UR, the turbulent field is also homogeneous in the axial direction. 
Thus, the mean velocity yields
\begin{align}
    \bm{U} = (U_r,U_\theta,U_z) = (0, U_\theta(r),0).
    \label{eq:meanvelocitycondition}
\end{align}
Hereafter, we consider the homogeneity of turbulent fields in both the azimuthal and axial directions. 
Therefore, the RANS equation for the azimuthal velocity yields
\begin{align}
    - \frac{1}{r^2} \frac{\d}{\d r} (r^2 R_{r\theta}) + \nu \frac{1}{r^2} \frac{\d}{\d r} \left[ r^3 \frac{\d}{\d r} \left( \frac{U_\theta}{r} \right) \right] = 0.
    \label{eq:ranseq}
\end{align}
The boundary conditions at the inner and outer cylinders are
\begin{gather}
    U_\theta (r=r_\mathrm{in}) = r_\mathrm{in} \omega_\mathrm{in} = U_\mathrm{in}, \ \ 
    U_\theta (r=r_\mathrm{out}) = r_\mathrm{out} \omega_\mathrm{out} = U_\mathrm{out}, 
    \nonumber \\
    R_{r\theta}(r=r_\mathrm{in}) = R_{r\theta}(r=r_\mathrm{out}) = 0,
    \label{eq:velocityboundaryconditions}
\end{gather}
where $\omega_\mathrm{in}$ and $\omega_\mathrm{out}$ denote the angular velocities of the inner and outer cylinders, respectively. 
In addition, $r_\mathrm{in}$ and $r_\mathrm{out}$ are the radii of the inner and outer cylinders, respectively. 
The parameters characterizing the TC flow are the Reynolds number $\Rey_\mathrm{in}$ based on the inner-cylinder velocity and gap width, radius ratio $\eta$, and angular velocity ratio $a$, which are defined as
\begin{align}
    \Rey_\mathrm{in} = \frac{U_\mathrm{in} d}{\nu}, \ \ 
    \eta = \frac{r_\mathrm{in}}{r_\mathrm{out}}, \ \ 
    a = - \frac{\omega_\mathrm{out}}{\omega_\mathrm{in}},
\end{align}
where $d = r_\mathrm{out} - r_\mathrm{in}$ is the gap width.
The flow regime of TC flows is often classified by the Taylor number $\Tay$, which is defined as
\begin{align}
    \Tay = \frac{(1+\eta)^6}{64\eta^4} (1+a)^2 \Rey_\mathrm{in}^2.
\end{align}
The details of the derivation and physical meaning of this Taylor number are provided by \citet{eckhardtetal2007}.
The UR of the TC turbulence is realized for the high-$\Tay$ regime:~$\Tay\gtrsim 10^9$ \citep{grossmannetal2016}.

Integrating Eq.~(\ref{eq:ranseq}) from the inner cylinder to $r$ provides a mean shear stress balance:
\begin{align}
    -r^2 R_{r\theta} + \nu r^3 \frac{\d}{\d r} \left( \frac{U_\theta}{r} \right) + r_\mathrm{in}^2 u_\tau^2 = 0,
    \label{eq:shearstressbalance}
\end{align}
where $u_\tau \left[= \sqrt{- \nu \left. r \d/\d r (U_\theta/r)\right|_{r=r_\mathrm{in}}}~\right]$ denotes the friction velocity. 
This conservation law was denoted as an angular velocity current $J^\omega$ by \citet{eckhardtetal2007}:
\begin{align}
    J^\omega = r^3 \langle u_r' \omega' \rangle - \nu r^3 \frac{\d \langle \omega \rangle}{\d r}
    = r^2 R_{r\theta} - \nu r^2 \frac{\d}{\d r} \left( \frac{U_\theta}{r} \right)
    = r_\mathrm{in} u_\tau^2 = \text{const.},
\end{align}
where $\omega = u_\theta/r$.
The angular velocity current is a significant quantity defining the Nusselt number in the TC flows, which is the measure of the efficiency of angular velocity transport.
In the bulk region, the viscous term decreases; thus, the mean shear stress balance yields
\begin{align}
    \frac{R_{r\theta}}{u_\tau^2} \simeq \left(\frac{r}{r_\mathrm{in}} \right)^{-2},
    \label{eq:shearstressasympte}
\end{align}
regardless of the outer cylinder rotation. 
Therefore, the Reynolds shear stress must be finite, even in a turbulent TC flow with outer cylinder rotation. 
Note that it is not possible to derive the mean velocity profile solely from the exact relation for the turbulence transport such as $R_{r\theta}$ or $J^\omega$, without using any assumptions or models.
In this study, we employ the RANS models to obtain the velocity profile.
The mean shear stress balance given by Eq.~(\ref{eq:shearstressbalance}) also holds for the RANS simulations; thus, the asymptote in the bulk region given by Eq.~(\ref{eq:shearstressasympte}) holds regardless of the model expression of the Reynolds stress at high Reynolds numbers.

\subsection{Reynolds stress transport equations for TC turbulence}\label{sec:reynoldsstresstransport}

For the featureless UR of the TC turbulence, the transport equations for the Reynolds stress with nonzero production terms in the inertial frame are
\begin{subequations}
\begin{align}
    \frac{\D R_{rr}}{\D t}
    = - \frac{2 U_\theta R_{r\theta}}{r}
    & = P_{rr} - \varepsilon_{rr} + \Phi_{rr} + D_{rr}
    \label{eq:rrrbudget}, \\
    \frac{\D R_{\theta \theta}}{\D t}
    = \frac{2 U_\theta R_{r\theta}}{r}
    & = P_{\theta\theta} - \varepsilon_{\theta\theta} + \Phi_{\theta\theta} + D_{\theta\theta}
    \label{eq:rttbudget}, \\
    \frac{\D R_{r \theta}}{\D t}
    = \frac{U_\theta (R_{rr}-R_{\theta\theta})}{r}
    & = P_{r\theta} - \varepsilon_{r\theta} + \Phi_{r\theta} + D_{r\theta}.
    \label{eq:rrtbudget}
\end{align}
\end{subequations}
The terms on the right-hand side are defined in the inertial frame as follows:
\begin{subequations}
\begin{align}
    P_{ij} & = - R_{i\ell} \nabla_\ell U_j - R_{j\ell} \nabla_\ell U_i
    \label{eq:production}, \\
    \varepsilon_{ij} & = 2\nu \langle (\nabla_\ell u_i') (\nabla_\ell u_j') \rangle
    \label{eq:dissipation}, \\
    \Phi_{ij} & = 2 \langle p' s_{ij}' \rangle
    \label{eq:pressurestrain}, \\
    D_{ij} & = - \nabla_\ell ( \langle u_i' u_j' u_\ell' \rangle 
    + \langle p' u_i' \rangle \delta_{j\ell} + \langle p' u_j' \rangle \delta_{i\ell}
    - \nu \nabla_\ell R_{ij} ).
    \label{eq:totaldiffusion}
\end{align}
\end{subequations}
They are referred to as the production, dissipation, pressure--strain correlation, and total diffusion terms, respectively. 
The strain rate $s_{ij}$ in the inertial frame is defined as
\begin{align}
    s_{ij} = \frac{1}{2} (\nabla_j u_i + \nabla_i u_j).
\end{align}
Owing to the curvature, the production term for the wall-normal stress component $P_{rr}$ is nonzero. 
Namely, for the featureless UR of the TC flows, the production terms yield
\begin{subequations}
\begin{align}
    P_{rr} & = -2R_{r\theta} \nabla_\theta U_r 
    = 2R_{r\theta} \frac{U_\theta}{r}, \\
    P_{\theta \theta} & = -2R_{\theta r} \nabla_r U_\theta 
    = -2R_{r\theta} \frac{\d U_\theta}{\d r}, \\
    P_{r\theta} & = - R_{rr} \nabla_r U_\theta - R_{\theta \theta} \nabla_\theta U_r
    = -R_{rr} \frac{\d U_\theta}{\d r} + R_{\theta \theta} \frac{U_\theta}{r}.
\end{align}
\end{subequations}

\subsection{Implicit algebraic model incorporating convection effects}\label{sec:modelling}

To verify the effect of flow curvature on the Reynolds stress in TC flows, we derive an algebraic Reynolds stress model (ARSM) expression based on the Reynolds stress transport equations (\ref{eq:rrrbudget})--(\ref{eq:rrtbudget}). 
According to primitive modeling by \citet{pope1975}, we also assume that the total diffusion term given by Eq.~(\ref{eq:totaldiffusion}) is negligible. 
In addition, we use the model for dissipation given by Eq.~(\ref{eq:dissipation}) and pressure--strain correlation given by Eq.~(\ref{eq:pressurestrain}), which are proposed by \citet{lrr1975}; namely, they are modeled as
\begin{subequations}
\begin{gather}
    \varepsilon_{ij} = \frac{2}{3} \varepsilon \delta_{ij},
    \label{eq:dissipationmodel} \\
    \Phi_{ij} = - C_\mathrm{S} \varepsilon b_{ij} 
    + C_\mathrm{R1} K S_{ij} 
    + C_\mathrm{R2} K \left[ S_{i\ell} b_{\ell j} + S_{j\ell} b_{\ell i} \right]_\mathrm{tl}
    + C_\mathrm{R3} K \left( W^\mathrm{A}_{i \ell} b_{\ell j} + W^\mathrm{A}_{j \ell} b_{\ell i} \right),
    \label{eq:pressurestrainmodel}
\end{gather}
\end{subequations}
where $C_\mathrm{S}$, $C_\mathrm{R1}$, $C_\mathrm{R2}$, and $C_\mathrm{R3}$ are the model constants, and $[A_{ij}]_\mathrm{tl} = A_{ij} - A_{\ell \ell} \delta_{ij}/3$. 
Note that we do not assume the weak-equilibrium condition, in contrast to \citet{pope1975}. 
Under these conditions, the Reynolds stress transport equations in the inertial frame yield
\begin{align}
    \frac{\D b_{ij}}{\D t} & = 
    - \left(C_\mathrm{S} - 1 + \frac{P^K}{\varepsilon} \right)\frac{\varepsilon}{K} b_{ij} 
    - \left(\frac{4}{3} - C_\mathrm{R1} \right) S_{ij}
    \nonumber \\
    & \hspace{1em}
    - (1 - C_\mathrm{R2}) \left[ S_{i\ell} b_{\ell j} + S_{j\ell} b_{\ell i} \right]_\mathrm{tl}
    - (1 - C_\mathrm{R3}) \left( W_{i \ell} b_{\ell j} + W_{j \ell} b_{\ell i} \right),
    \label{eq:reynoldsstresstransportmodel}
\end{align}
where $P^K (= P_{ii}/2)$ denotes the production rate of the turbulent kinetic energy.
For the featureless UR of the TC turbulence, the leading components of the Reynolds stress yield the following matrix equation:
\begin{align}
    &
    \begin{bmatrix}
    \dfrac{\varepsilon}{gK} &
    0 & \dfrac{2}{3} C_2 S_{r\theta} + 2 C_3 W_{r \theta} - \dfrac{2U_\theta}{r} \\
    0 & \dfrac{\varepsilon}{gK} & \dfrac{2}{3} C_2S_{r\theta} - 2 C_3 W_{r \theta} + \dfrac{2U_\theta}{r}  \\
    C_2 S_{r\theta} - C_3 W_{r \theta} + \dfrac{U_\theta}{r} &
    C_2 S_{r\theta} + C_3 W_{r \theta} - \dfrac{U_\theta}{r} &
    \dfrac{\varepsilon}{gK}
    \end{bmatrix}
    \begin{bmatrix}
    b_{rr} \\ b_{\theta \theta} \\ b_{r\theta}
    \end{bmatrix}
    \nonumber \\
    & =
    - 2C_1 K
    \begin{bmatrix}
    0 \\ 0 \\ S_{r\theta}
    \end{bmatrix},
\end{align}
where
\begin{gather}
    S_{r\theta} = \frac{1}{2} r \frac{\d}{\d r} \left( \frac{U_\theta}{r} \right), \ \ 
    W_{r\theta} = - \frac{1}{2r} \frac{\d}{\d r} (rU_\theta),
    \nonumber \\
    C_1 = \frac{2}{3} - \frac{C_\mathrm{R1}}{2}, \ \ 
    C_2 = 1 - C_\mathrm{R2}, \ \ 
    C_3 = 1 - C_\mathrm{R3}, \ \
    g = \left( C_\mathrm{S} - 1 + \frac{P^K}{\varepsilon} \right)^{-1}.
    \label{eq:redefinedconstants}
\end{gather}
Thus, the solution for shear stress yields
\begin{align}
    R_{r\theta} (=Kb_{r\theta} ) = 
    - \frac{2C_1}{1 + 4 (g \tau)^2 [- C_2^2 S_{r\theta}^2/3 + (C_3 W_{r\theta} - U_\theta/r)^2]} g \tau K S_{r\theta},
    \label{eq:implicitarsm}
\end{align}
where $\tau = K/\varepsilon$. 
In this expression, $U_\theta/r$ in the denominator results from the convection term of the Reynolds shear stress. 
This form of the model is essentially the same as that proposed by \citet{wj2000} for the two-dimensional case with its extension to the curved flows \citep{wj2002}.
Strictly speaking, we should consider the $S_{ij}$ dependence of $P^K$ to derive a fully explicit ARSM \citep{girimaji1996}. 
We do not consider such sophistication to verify the convection effects on the Reynolds stress $U_\theta/r$ through the denominator. 
This representation of the convection effects is similar to the history effect in a swirling flow in a straight pipe proposed by \citet{hamba2017}.
They are essentially the same because they result from the Lagrangian derivative of the Reynolds stress.
Thus, part of the flow curvature effect can be expressed by considering the history effect of the Reynolds stress.
In several ARSMs, the $S_{r\theta}^2$ part in the denominator of Eq.~(\ref{eq:implicitarsm}) is often neglected because $C_3 > C_2 = 1-C_\mathrm{R2} \simeq 0$ is suggested by second-order or Reynolds stress transport models (e.g., Refs.~\citep{taulbee1992,wj2000}).
In this study, we first verify the performance of the model based on Eq.~(\ref{eq:implicitarsm}) to extract the essence of representing the curvature effects. 
Note that generalizing the model provided by Eq.~(\ref{eq:implicitarsm}) to a covariant representation is not necessarily straightforward.
The generalization of the model, specifically for incorporating the convection effects, is discussed using a covariant time derivative in Sec.~\ref{sec:discussion}.

\section{Verification of RANS model compared with experiments}\label{sec:verification}
In this section, we compare the performances of ARSMs with the experimental results to confirm which term significantly contributes to the Reynolds stress in the budget. 

\subsection{Experimental setup}\label{sec:experimentalsetup}
We utilized a very large facility to realize high-Reynolds-number TC turbulence, also used in our previous experiments \citep{horimoto2025}.
The TC facility consists of an aluminum inner and an acrylic outer cylinder whose radii are $r_\mathrm{in}=150$ mm and $r_\mathrm{out}=205$ mm, respectively.
The height of the annulus region between these cylinders is $L=990$ mm. 
Thus, the dimensionless geometric parameters in the present experiments are the radius ratio of $\eta= 0.732$ and the aspect ratio of $\varGamma=L/d=18$.
The angular velocities of the two cylinders, $\omega_\mathrm{in}$ and $\omega_\mathrm{out}$, are independently controlled by two stepper motors.
The working fluid was degassed water.
The examined parameter regimes are $\Rey_\mathrm{in} = O(10^4)$ and $-0.5\leq a\leq 0.1$, which result in a $\Tay$ regime of $O(10^9)$--$O(10^{10})$, as shown in Table~\ref{tab:ExperimentalParameters}.
Note that the present setup achieves a sufficiently high $\Tay$ for the UR: $\Tay \gtrsim O(10^9)$ and, based on the flow regime diagrams for $\eta=0.714$ explored by \citet{ostillamonicoetal2014c}, the futureless UR for the cases of $a\leq0$ (see also Ref.~\cite{grossmannetal2016}).
In our preliminary flow visualization using light-reflective flakes, we observed the absence of inflow and outflow generated by the Taylor rolls in co-rotation conditions.
For further evidence, it is necessary to measure three-dimensional mean flow structure, e.g., by traversing the measurement plane as in \citet{ezetaetal2020}, which is impossible with the present apparatus.

\begin{table}
%%\begin{center}
%%\def~{\hphantom{0}}
\centering
\caption{Control parameters examined by the present experimental facility with
    $\eta=0.732$ and $\varGamma=18$. $\Rey_\mathrm{out}$ is the outer cylinder
    Reynolds number:~$\Rey_\mathrm{out}=U_\mathrm{out}d/\nu$. The shear Reynolds number $\Rey_\mathrm{s}$ is given by Eq.~(\ref{def:Re_s}).}
\begin{ruledtabular}
\begin{tabular}{ccccc}
  $a$ & $\Rey_\mathrm{in}$ & $\Rey_\mathrm{out}$ & $\Tay$ & $\Rey_\mathrm{s}$ \\[3pt] \hline
  $-0.5$ & $8.5\times 10^4$ & $\mspace{11.5mu}5.8\times 10^4$ & $2.7\times 10^9$ & $4.9\times10^4$ \\
  ~$-0.33$ & $4.1\times 10^4$ & $\mspace{11.5mu}1.9\times 10^4$ & $1.1\times 10^9$ & $3.1\times10^4$ \\
  ~$-0.33$ & $6.2\times 10^4$ & $\mspace{11.5mu}2.8\times 10^4$ & $2.5\times 10^9$ & $4.8\times10^4$ \\
  ~$-0.33$ & $8.5\times 10^4$ & $\mspace{11.5mu}3.9\times 10^4$ & $4.7\times 10^9$ & $6.5\times10^4$ \\
  $-0.1$ & $8.5\times 10^4$ & $\mspace{11.5mu}1.2\times 10^4$ & $8.6\times 10^9$ & $8.8\times10^4$ \\
  $\mspace{-2.5mu}0$ & $4.1\times 10^4$ & 0 & $2.4\times 10^9$ & $4.7\times10^4$ \\
  $\mspace{-2.5mu}0$ & $6.2\times 10^4$ & 0 & $5.6\times 10^9$ & $7.2\times10^4$ \\
  $\mspace{-2.5mu}0$ & $8.5\times 10^4$ & 0 & $\mspace{7.5mu}1.1\times 10^{10}$ & $9.8\times10^4$ \\
  $\mspace{11.5mu}0.1$ & $8.5\times 10^4$ & $-1.2\times 10^4$ & $\mspace{7.5mu}1.3\times 10^{10}$ & $1.1\times10^5$ 
\end{tabular}
\end{ruledtabular}
\label{tab:ExperimentalParameters}
%%\end{center}
\end{table}

We measured the turbulent velocity field in the $r$--$\theta$ plane at half-height of the gap.
As tracer particles, we dispersed silver-coated hollow glass particles (S-HGS-10, Dantec Dynamics. Diameter $D=10$~\textmu m, apparent density $\rho_p=1010$~kg/m$^3$), which were illuminated by a laser sheet (DPRLu-5W, Japan Laser. Wavelength 640~nm, thickness 1~mm).
These particles provided high traceability for flow;~the Stokes number, $\Sto=\rho_p D^2/18\rho\nu T$, of the particles is sufficiently low, $\Sto\leq O(10^{-4})$, for all examined parameters.
Here, $\rho$ is the density of water and $T=d/(U_\mathrm{in} - U_\mathrm{out})$ is the time scale of the bulk flow, respectively.
Particle motion was recorded by a high-speed camera (AX-50, Photoron) with a micro lens (Micro-Nikkor 105mm f/2.8, Nikon) set above the apparatus in the laboratory frame.
The resolution of the camera was 1024~pixels$\times$1024~pixels, corresponding to a field of view of about 75.4~mm$\times$75.4~mm, including the side surfaces of the cylinders.
Frame rates were adjusted based on $\Rey_\mathrm{in}$ (i.e., the higher wall speed $U_\mathrm{in}$):~1000~fps for $\Rey_\mathrm{in}=4.1\times10^4$ and 2000~fps for $\Rey_\mathrm{in}=6.2\times10^4$ and $\Rey_\mathrm{in}=8.5\times10^4$.
The dimensionless measurement time was at least 8.1$T$ for $a=0.1$ and $\Rey_\mathrm{in}=8.5\times10^4$, which is limited by the memory capacity of the camera.
This implies that the measurement time is several times larger than the time scale of the largest-scale motion in the flow.
We confirmed the reproducibility of the following results by conducting experiments twice for all examined parameters.
In addition, the mean angular velocity profiles of our experiments agreed well with the experiments by \citet{vangilsetal2012}, where $\eta = 0.716$ for their apparatus (figure omitted).

Regarding velocimetry, we used a hybrid algorithm combining particle image velocimetry (PIV) and particle tracking velocimetry (PTV), based on an idea similar to previously proposed algorithms \citep{stitou2001,Sussetetal2006}.
First, we employed a direct cross-correlation method like usual PIV, using an interrogation area centered at each detected particle position, to predict particle displacement.
Next, we detected the particles around the predicted position in the second frame.
Among them, we determined the particle whose surrounding particle pattern, within an area smaller than that for the above prediction, provided the maximum correlation with that in the first frame.
To avoid erroneous matching, we set a criterion for this correlation and applied the inverse particle matching (i.e., from the second to the first frames);~only particle pairs passing both checks were used to calculate velocity.
Note that these procedures detected the displacements of individual particles in the first frame like PTV, providing finer spatial resolution compared to those obtained only by the PIV method.
Finally, we removed error vectors with a threshold based on the first and third quantiles of the two velocity components.
The number of obtained velocity vectors was about 5000 in each frame.
Here, note that these vectors were spatially irregular because we tracked the dispersed tracer particles.
Therefore, the gap width was divided into 100 thin annular regions with a constant width $d/100$, and the mean velocity $U_\theta(r)$ was calculated by averaging the obtained velocity vectors over the azimuthal and time directions in each annulus.

Here, we compare the temporal and spatial resolutions in our measurements with the smallest scales in the bulk turbulence.
To this end, we employed the shear Reynolds number \citep{dubrulleetal2005}, defined as
\begin{align}
    \Rey_\mathrm{s}=\frac{2}{1+\eta}\left|\eta\Rey_\mathrm{out}-\Rey_\mathrm{in}\right|,
    \label{def:Re_s}
\end{align}
(see also Table~\ref{tab:ExperimentalParameters} for $\Rey_\mathrm{s}$ of the present experiments).
The acquisition period of the velocity field, the inverse of the frame rates, was of the same order as the Kolmogorov time-scale $t_\mathrm{K}$ estimated as $t_\mathrm{K}=T\Rey_\mathrm{s}^{-1/2}$;~about $5t_\mathrm{K}$ at longest for the highest-$\Rey_\mathrm{s}$ case of $\Rey_\mathrm{s}=1.1\times10^5$ (the only counter-rotating case).
The average distance between the particles with the obtained velocity vector was about 0.5~mm, which was about $O(10)$ times larger than the Kolmogorov length scale $\eta_\mathrm{K}$, estimated as $\eta_\mathrm{K}=d\Rey_\mathrm{s}^{-3/4}$;~50 times for the highest-$\Rey_\mathrm{s}$ case at the largest.
However, these resolutions were sufficient to measure the nearly constant profile of mean angular momentum in the bulk region (recall that we confirmed its reproducibility).

\subsection{Numerical setup}\label{sec:numericalsetup}

We performed the $K$--$\varepsilon$ RANS model. 
The baseline model is the conventional linear eddy-viscosity model proposed by \citet{akn1994}, which allows us to use the no-slip condition owing to the damping functions for wall-bounded flows. 
The governing equations are the RANS equation for the mean azimuthal velocity given by Eq.~(\ref{eq:ranseq}) and transport equations for the turbulent kinetic energy $K$ and its dissipation rate $\varepsilon$:
\begin{align}
    \frac{\D K}{\D t} & = 
    - 2R_{r\theta} S_{r\theta}
    - \varepsilon 
    + \frac{1}{r} \frac{\d}{\d r} \left[ r \left( \frac{\nu_\mathrm{T}}{\sigma_K} + \nu \right) \frac{\d K}{\d r} \right] = 0, 
    \label{eq:keq} \\
    \frac{\D \varepsilon}{\D t} & = 
    - 2C_{\varepsilon 1} \frac{\varepsilon}{K} R_{r\theta} S_{r\theta}
    - C_{\varepsilon 2} f_\varepsilon \frac{\varepsilon^2}{K} 
    + \frac{1}{r} \frac{\d}{\d r} \left[ r \left( \frac{\nu_\mathrm{T}}{\sigma_\varepsilon} + \nu \right) \frac{\d \varepsilon}{\d r} \right] = 0,
    \label{eq:epsiloneq}
\end{align}
where $\nu_\mathrm{T}$ is the eddy viscosity, which is defined as
\begin{align}
    \nu_\mathrm{T} = C_\nu f_\nu \frac{K^2}{\varepsilon}.
    \label{eq:eddyviscosity}
\end{align}
$f_\nu$ and $f_\varepsilon$ are damping functions defined as
\begin{align}
    f_\nu & = \left\{ 1 -\mathrm{exp} \left[ - \frac{y}{a_1\eta_\mathrm{K}} \right] \right\}^2
    \left\{ 1 + \frac{a_2}{\Rey_\mathrm{T}^{3/4}} \mathrm{exp} \left[ - \left( \frac{\Rey_\mathrm{T}}{a_3} \right)^2 \right] \right\}, \\
    f_\varepsilon & = \left\{ 1 -\mathrm{exp} \left[ - \frac{y}{a_{\varepsilon 1} \eta_\mathrm{K}} \right] \right\}^2
    \left\{ 1 - a_{\varepsilon 2} \mathrm{exp} \left[ - \left( \frac{\Rey_\mathrm{T}}{a_{\varepsilon 3}} \right)^2 \right] \right\},
\end{align}
where $y$, $\eta_\mathrm{K} [= (\nu^3/\varepsilon)^{1/4}]$, and $\Rey_\mathrm{T} [= K^2/(\nu \varepsilon)]$ denote the distance from the nearest wall, Kolmogorov length scale, and turbulent Reynolds number, respectively. 
Note again that we consider the featureless UR; thus, the statistics depend only on the radial position $r$. 
The model parameters were set as follows:
\begin{gather}
    C_\nu = 0.09, \
    C_{\varepsilon1} = 1.5, \
    C_{\varepsilon2} = 1.9, \
    \sigma_K = 1.4, \
    \sigma_\varepsilon = 1.4, 
    \nonumber \\
    a_1 = 14, \
    a_2 = 5, \
    a_3 = 200, \
    a_{\varepsilon1} = 3.1, \
    a_{\varepsilon2} = 0.3, \
    a_{\varepsilon3} = 6.5.
\end{gather}
The velocity and length were normalized by the inner-cylinder velocity $U_\mathrm{in} (=r_\mathrm{in} \omega_\mathrm{in})$ and gap width between the cylinders $d$. 
The boundary conditions were Eq.~(\ref{eq:velocityboundaryconditions}) and
\begin{gather}
    K(r=r_\mathrm{in}) = K(r=r_\mathrm{out}) = 0, 
    \nonumber \\ 
    \varepsilon(r=r_\mathrm{in}) = \nu \frac{1}{r} \frac{\d}{\d r} \left( r \frac{\d K}{\d r} \right)(r=r_\mathrm{in}), \ \
    \varepsilon(r=r_\mathrm{out}) = \nu \frac{1}{r} \frac{\d}{\d r} \left( r \frac{\d K}{\d r} \right)(r=r_\mathrm{out}).
\end{gather}

For the original linear eddy-viscosity model proposed by \citet{akn1994}, the Reynolds shear stress is given by
\begin{align}
    R_{r\theta} = - 2\nu_\mathrm{T} S_{r\theta}.
    \label{eq:lineareddyviscosity}
\end{align}
Hereafter, we refer to the model using Eqs.~(\ref{eq:ranseq}), (\ref{eq:keq}), (\ref{eq:epsiloneq}), (\ref{eq:eddyviscosity}), and (\ref{eq:lineareddyviscosity}) as the AKN model. 
To observe the effect of the flow curvature, we employed the following algebraic model of the Reynolds shear stress according to Eq.~(\ref{eq:implicitarsm}):
\begin{align}
    R_{r\theta} = - \frac{2C_1}{1 + 4 \tau_\mathrm{T}^2 (C_3 W_{r\theta} - C_r U_\theta/r)^2} \tau_\mathrm{T} K S_{r\theta}, \
    \tau_\mathrm{T} = C_\tau f_\nu \frac{K}{\varepsilon}.
    \label{eq:ccarsm}
\end{align}
We refer to this model as the curvature-corrected ARSM (ccARSM) because this model is essentially similar to that proposed by \citet{wj2002}. 
The effects of flow curvature emerge via the vorticity $W_{r\theta}$ or the convection originated term $U_\theta/r$.
$C_\tau$ corresponds to $C_\mathrm{S}^{-1}$ for the production--dissipation equilibrium condition $P^K/\varepsilon = 1$ in the denominator. 
Compared with Eq.~(\ref{eq:implicitarsm}), we removed the $S_{r\theta}^2$ term from the denominator for the ccARSM, as is often assumed in several ARSMs (e.g., Refs.~\cite{taulbee1992,wj2000}).
We also examined the model with the strain rate in the denominator and confirmed that it was ineffective in predicting TC flows (see Appendix~\ref{sec:modelparameter}). 
Note that $C_r$, which is unity in Eq.~(\ref{eq:implicitarsm}), is an artificially introduced parameter to examine the convection effects via $U_\theta/r$. 
In other words, the convection effects can be switched off by setting $C_r = 0$, which corresponds to the standard ARSM proposed by \citet{pope1975}, although the strain rate was removed from the denominator of the Reynolds shear stress. 
Hereafter, we fix $C_1 = C_\nu/C_\tau$ so that the ccARSM reduces to the AKN model when $C_3=C_r=0$.
In addition, we fix $C_\tau = 1/3.9$ so that the ccARSM provides a good prediction for various flow parameters. 
In this study, we focus on the existence of contributions of $W_{r\theta}$ or $U_\theta/r$ to the Reynolds stress in the denominator; therefore, we set $C_3$ and $C_r$ to $0$ or $1$.

\subsection{Results}\label{sec:results}

\subsubsection{Experimental results}\label{sec:resultsexp}

\begin{figure}
    \centering
    \includegraphics[width=0.5\linewidth]{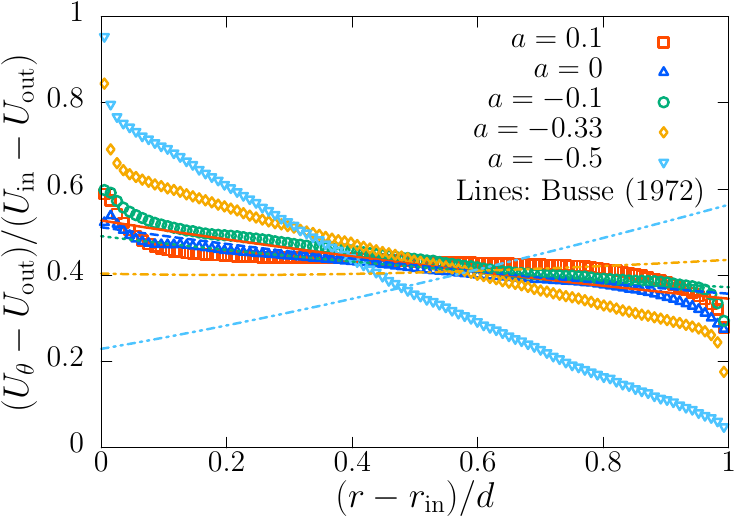}
    \caption{Mean velocity profiles of the experiments at $\Rey_\mathrm{in}=8.5\times 10^4$ for various angular velocity ratios $a$.
    The lines are those theoretically proposed by \citet{busse1972}: solid red, $a = 0.1$; dash blue, $a = 0$; dot green, $a = -0.1$; dash-dot yellow, $a = -0.33$; dash-dot-dot skyblue, $a = -0.5$.
    }
    \label{fig:meanexp}
\end{figure}

Figure~\ref{fig:meanexp} depicts the mean velocity profiles in the experiments for various angular velocity ratios $a$. 
The mean velocity is normalized such that it yields values of $1$ and $0$ at the inner and outer cylinders, respectively. 
In this normalization, the mean velocity gradient becomes steeper as $a$ decreases. 
The TC flow is Rayleigh stable for $r_\mathrm{in}^2 \omega_\mathrm{in} < r_\mathrm{out}^2 \omega_\mathrm{out}$ or $-a > \eta^2$ for $\omega_\mathrm{in} > 0$ \citep{rayleigh1917}.
For our experimental apparatus, the critical line is $a = -0.536$. 
Therefore, $a=-0.5$ is close to the stability line; thus, the mean velocity profile looks similar to the laminar profile.
\citet{busse1972} proposed a velocity profile corresponding to the asymptotic limit of the variational problem for the upper bound on the transport of angular momentum.
It yielded
\begin{align}
    \frac{U_\theta}{U_\mathrm{in}} 
    = \frac{\eta(1+a)}{4(1-\eta) (1-\eta^2)} \frac{d}{r}
    + \frac{\eta^3 (1-2\eta^2) - a \eta (2-\eta^2)}{2 (1-\eta) (1-\eta^4)} \frac{r}{d}.
\end{align}
We also depict the profiles for each $a$ in Fig.~\ref{fig:meanexp}.
The slopes clearly differ from the experimental results in the bulk region, except for $a=0$.
This discrepancy was already pointed out by \citet{vangilsetal2012}.
Thus, the mean velocity profiles are determined via a different mechanism from the upper bound on the angular momentum transport.

\begin{figure}[t]
    \centering
    \begin{minipage}[t]{.48\linewidth}
        \centering
        \includegraphics[width=\linewidth]{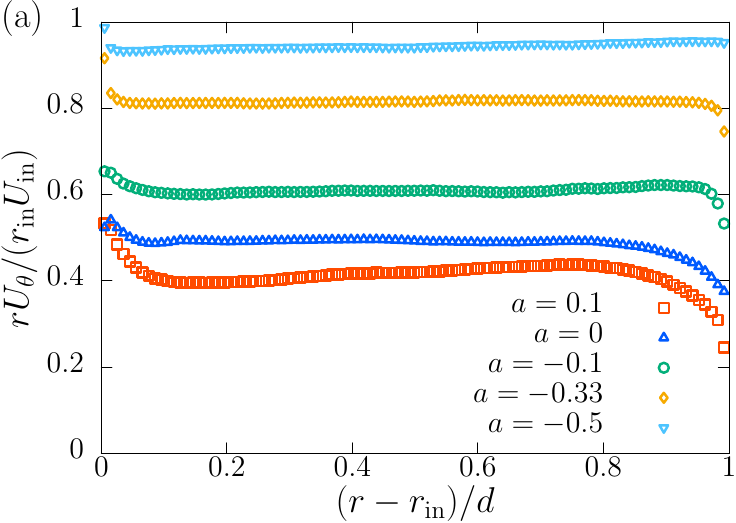}
    \end{minipage}
    \hspace{.01\linewidth}
    \begin{minipage}[t]{.48\linewidth}
        \centering
        \includegraphics[width=\linewidth]{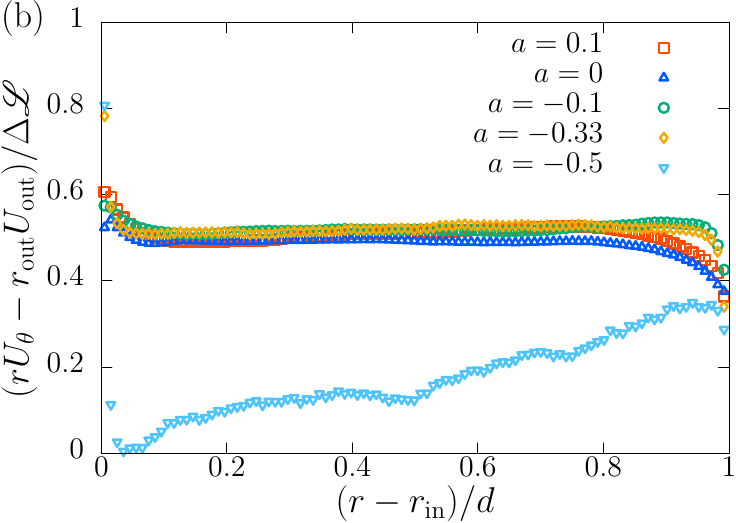}
    \end{minipage}
    \caption{Mean angular momentum profiles of the experiments at $\Rey_\mathrm{in}=8.5\times 10^4$ for various $a$, normalized by (a) the angular momentum of the inner cylinder and (b) the angular momentum difference between the inner and outer cylinders.}
    \label{fig:angnularexp}
\end{figure}

Figure~\ref{fig:angnularexp}(a) depicts the mean angular momentum profiles normalized by the angular momentum of the inner cylinder.
The nearly constant angular momentum seems universal in the bulk region $0.1 < (r-r_\mathrm{in})/d < 0.9$ for $a < 0.1$. 
In this normalization, the constant value of the angular momentum increases as $a$ decreases.
This trend is natural because the angular momentum will be bounded by $-a/\eta^2 \le r U_\theta/(r_\mathrm{in} U_\mathrm{in}) \le 1$ if its profile is almost monotonic.
For $a=0.1$, the angular momentum exhibits a shallow positive gradient.
This may be caused by the Taylor rolls.
To verify this, we must perform averaging over different heights in the future.

\citet{brauckmannetal2016} numerically showed that the mean angular momentum profiles collapse to $0.5$ in a bulk region for weakly counter-rotating and co-rotating cases when they are normalized by the angular momentum difference between the inner and outer cylinders and adjusted to $1$ and $0$ at the inner and outer cylinders, respectively. 
Figure~\ref{fig:angnularexp}(b) depicts the mean angular momentum profiles normalized by the angular momentum difference between the inner and outer cylinders: $\Delta \mathscr{L} = r_\mathrm{in} U_\mathrm{in} - r_\mathrm{out} U_\mathrm{out} = r_\mathrm{in}^2 \omega_\mathrm{in} - r_\mathrm{out}^2\omega_\mathrm{out}$. 
In addition, they are adjusted to $1$ and $0$ at the inner and outer cylinders, respectively. 
Our results also collapse to $0.5$ except for $a=-0.5$.
For our apparatus, $a=-0.5$ is close to the Rayleigh stable line: $a=-\eta^2 = -0.536$. 
When $a$ is close to the Rayleigh stable line, the denominator $\Delta \mathscr{L}$ becomes small; thus, the result is much more susceptible to errors. 
Furthermore, the flow may no longer be in the UR owing to the stabilization effect. 
The outlier for $a=-0.5$ may be caused by this flow property or by related difficulties in velocimetry.

\subsubsection{Performance of ccARSM}\label{sec:resultsmodel}

\begin{figure}[t]
    \centering
    \begin{minipage}[t]{.48\linewidth}
        \centering
        \includegraphics[width=\linewidth]{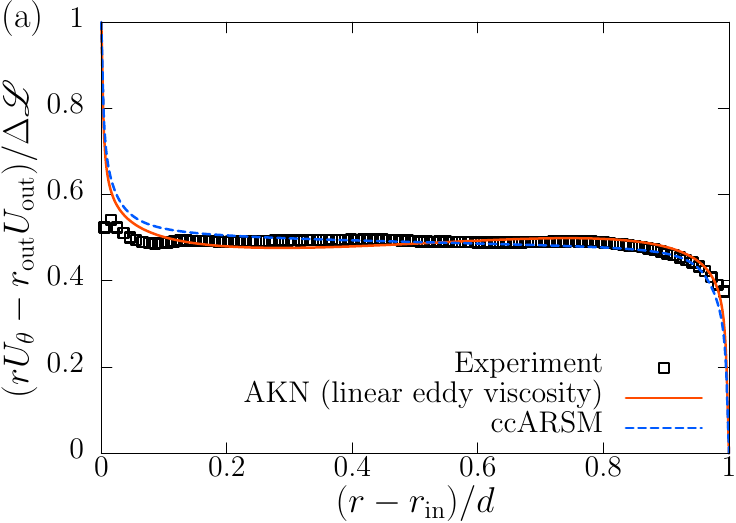}
    \end{minipage}
    \hspace{.01\linewidth}
    \begin{minipage}[t]{.48\linewidth}
        \centering
        \includegraphics[width=\linewidth]{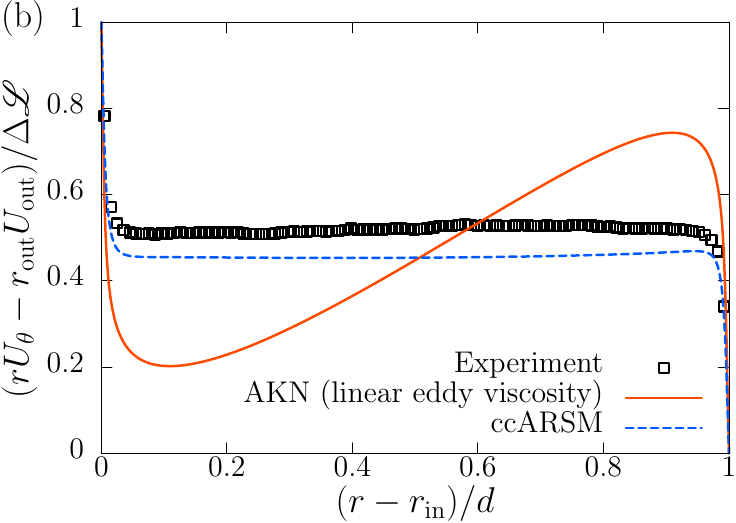}
    \end{minipage}
    \caption{Mean angular momentum profiles of AKN and ccARSM compared with those of experiments at $\Rey_\mathrm{in}=8.5\times 10^4$ for (a) $a=0$ and (b) $a=-0.33$.}
    \label{fig:modelangularmomentum}
\end{figure}

Figure~\ref{fig:modelangularmomentum} depicts the mean angular momentum profiles of the RANS models compared with the experiments at $\Rey_\mathrm{in} = 8.5\times 10^4$. 
Hereafter, we adopt the normalization by the angular momentum difference between the inner and outer cylinders.
As seen in Fig.~\ref{fig:modelangularmomentum}(a), both the AKN (linear eddy-viscosity) model and ccARSM predict the mean angular momentum for $a=0$. 
However, Fig.~\ref{fig:modelangularmomentum}(b) shows that the AKN model fails to predict the nearly constant mean angular momentum for $a=-0.33$. 
Therefore, the conventional linear eddy-viscosity model is irrelevant for predicting TC flows. 
In contrast, the ccARSM fairly succeeds in predicting a nearly constant mean angular momentum for both $a=0$ and $a=-0.33$. 
The emergence of a nearly constant mean angular momentum is independent of the Reynolds and Taylor numbers for both experiments and ccARSM for $\Tay \gtrsim O(10^9)$ (see Appendix~\ref{sec:reynoldsnumberdependence}).

\begin{figure}
    \centering
    \includegraphics[width=0.5\linewidth]{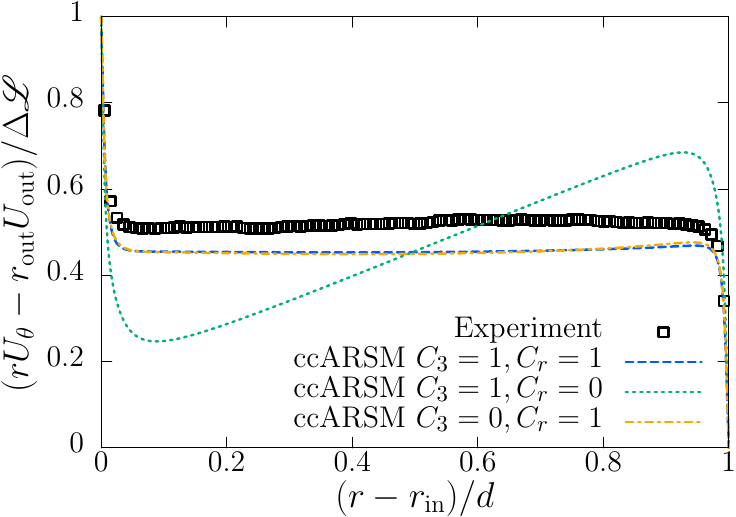}
    \caption{Mean angular momentum profiles for several sets of model parameters of ccARSM at $\Rey_\mathrm{in}=8.5\times 10^4$ for $a=-0.33$.}
    \label{fig:angularmodeldependence}
\end{figure}

Figure~\ref{fig:angularmodeldependence} depicts the mean angular momentum for several sets of model parameters of ccARSM for $a=-0.33$. 
When the convection effects are removed by setting $C_r=0$, the model fails to predict an experimental profile, similarly to the AKN model for $a=-0.33$ depicted in Fig.~\ref{fig:modelangularmomentum}(b).
In contrast, even when we set $C_3=0$, the ccARSM fairly reproduced the nearly constant mean angular momentum, similarly to the case of $C_3=C_r=1$. 
Therefore, we can conclude that the $U_\theta/r$-related part in the denominator of Eq.~(\ref{eq:ccarsm}) is essential for expressing the curvature effects in TC flows.
The difference between the present ccARSM and that proposed by \citet{wj2002} is only the value of constant parameters and near-wall treatments.
We infer that the explicit ARSM proposed by \citet{wj2002} also predicts the nearly constant mean angular momentum as it has the essence, $U_\theta/r$-related part.
The fine-tuning of constant parameters is out of the scope of this study.

\section{Discussion}\label{sec:discussion}

\subsection{Failure of linear eddy-viscosity model}\label{sec:failure}

As observed in Fig.~\ref{fig:modelangularmomentum}, the linear eddy-viscosity model (AKN) failed to predict the nearly constant mean angular momentum for $a = -0.33$.
The cause of this failure arises from the physical basis of the standard $K$--$\varepsilon$ model, which has been developed for predicting parallel shear flows.

According to our results of the nearly constant mean angular momentum in Fig.~\ref{fig:angnularexp}, we have (see also Ref.~\cite{brauckmannetal2016})
\begin{align}
    \frac{rU_\theta - r_\mathrm{out} U_\mathrm{out}}{\Delta \mathscr{L}} \simeq \frac{1}{2}
    \ \iff \ 
    r U_\theta \simeq \frac{1}{2} (r_\mathrm{in} U_\mathrm{in} + r_\mathrm{out} U_\mathrm{out}).
    \label{eq:constantangularmomentumre}
\end{align}
Combined with Eq.~(\ref{eq:shearstressasympte}), the production term for turbulent kinetic energy $P^K$ in the featurless UR for TC flows yields
\begin{align}
    P^K = - R_{r\theta}~r \frac{\d}{\d r} \left( \frac{U_\theta}{r} \right)
    \simeq \frac{r_\mathrm{in}^2}{r^2} u_\tau^2 \times \frac{1}{r^2}(r_\mathrm{in} U_\mathrm{in} + r_\mathrm{out} U_\mathrm{out})
    = \frac{r_\mathrm{in}^4}{r^4} u_\tau^2 \omega_\mathrm{in} (1 - a/\eta^2).
    \label{eq:exactproductionscaling}
\end{align}
When the production balances the dissipation, $\varepsilon$ has the same scaling.
In contrast, the standard $K$--$\varepsilon$ model has been developed to be consistent with the following scaling in the logarithmic region near the wall for plane shear flows \citep{yoshizawabook,popebook}:
\begin{align}
    P^K \simeq \varepsilon \sim \frac{\text{(velocity scale)}^3}{\text{(length scale)}} \sim \frac{u_\tau^3}{y},
    \label{eq:standardkepsilonscaling}
\end{align}
where $y$ denotes the distance from the wall.

\begin{figure}[t]
    \centering
    \begin{minipage}[t]{.48\linewidth}
        \centering
        \includegraphics[width=\linewidth]{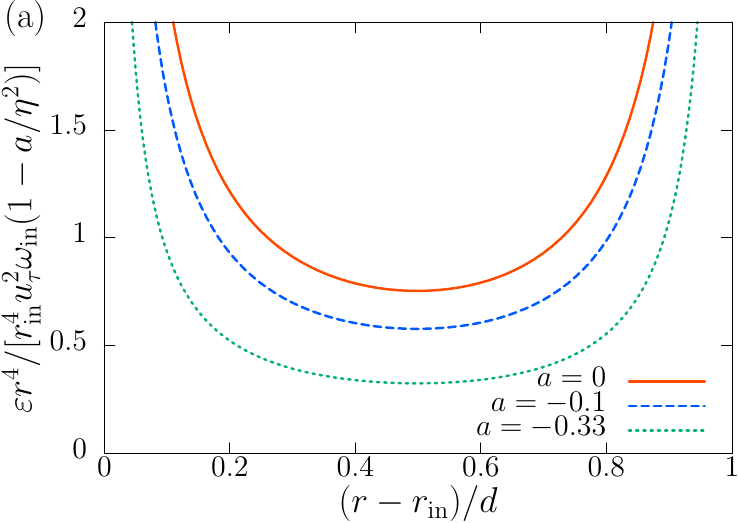}
    \end{minipage}
    \hspace{.01\linewidth}
    \begin{minipage}[t]{.48\linewidth}
        \centering
        \includegraphics[width=\linewidth]{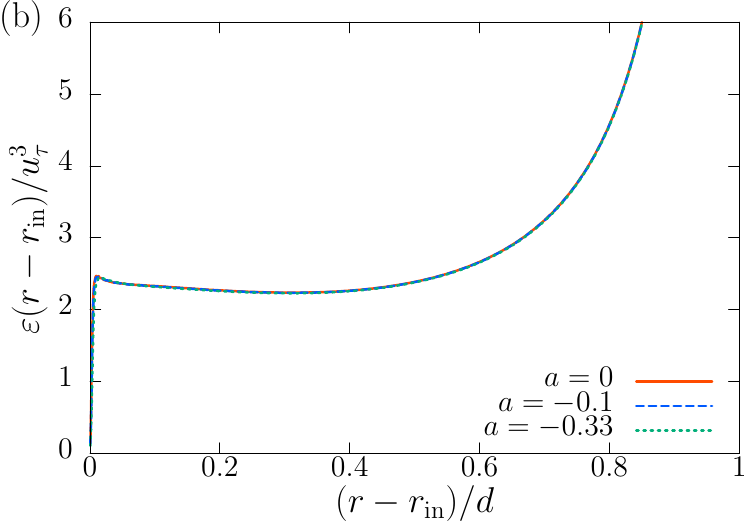}
    \end{minipage}
     \caption{Profiles of dissipation rate normalized by (a) Eq.~(\ref{eq:exactproductionscaling}) and (b) Eq.~(\ref{eq:standardkepsilonscaling}) with $y = r-r_\mathrm{in}$ for the AKN model.}
    \label{fig:epsilonscaling}
\end{figure}

Figure~\ref{fig:epsilonscaling} depicts the scaling of the dissipation rate for the AKN model.
As observed in Fig.~\ref{fig:epsilonscaling}(a), the results of the AKN model are inconsistent with the exact scaling for the featureless UR of the TC turbulence given by Eq.~(\ref{eq:exactproductionscaling}). 
In contrast, Fig.~\ref{fig:epsilonscaling}(b) shows that they are consistent with the scaling based on the log-law, given by Eq.~(\ref{eq:standardkepsilonscaling}) where $y = r-r_\mathrm{in}$.
This indicates that the standard $K$--$\varepsilon$ model with the linear eddy-viscosity expression does not represent the effects of flow curvature even though the RANS equations are discretized in cylindrical coordinates as written by Eq.~(\ref{eq:ranseq}).
Hence, the flow curvature effects must be properly implemented in the eddy viscosity to predict the nearly constant mean angular momentum.

\subsection{Flow curvature effects resulting from time history}\label{sec:genralization}

The success of the ccARSM in predicting the nearly constant mean angular momentum relies on the $U_\theta/r$-related part in the denominator of the Reynolds shear stress in Eq.~(\ref{eq:ccarsm}), which results from the convection of the Reynolds stress in Eqs.~(\ref{eq:rrrbudget})--(\ref{eq:rrtbudget}).
Therefore, the effect is considered a history effect of the Reynolds stress.
The history effect of the Reynolds stress is often essential in rotating flows \citep{hamba2006,hamba2017} and unsteady flows \citep{hd2008,ai2023}.
The history effect is related to the time derivatives.
The ARSM involving the time derivative of the strain rate has been discussed in several studies \citep{speziale1987,ac1991,taulbee1992,hd2008} and derived from statistical closure theories \citep{tsdia,ariki2019,ai2023}. 
\citet{ss1997} proposed a scalar measure of the curvature effects using the Lagrangian derivative of the strain rate. 
In this sense, incorporating the flow curvature effects using time derivative terms is physically plausible.
Furthermore, \citet{hamba2006} employed the Lagrangian time derivative transformed into a rotating frame of $S_{i\ell} W^\mathrm{A}_{\ell j} + S_{j\ell} W^\mathrm{A}_{\ell i}$ to elucidate the occurrence of nearly zero mean absolute vorticity profiles in the bulk of spanwise-rotating turbulent channel flows. 
Note that several studies have employed the upper-convected or Oldroyd time derivative instead of the Lagrangian derivative to achieve a covariant expression of ARSM \citep{speziale1987,hd2008,ariki2019,ai2023}.

Here, we consider the following modeled transport equations of the anisotropy tensor $b_{ij}$ with a linear relaxation model, which are formally the same as those proposed by \citet{lrr1975}:
\begin{align}
    \frac{\D b_{ij}}{\D t} = - \frac{b_{ij}}{\tau} + X_{ij},
    \label{eq:lagrangianrelaxation}
\end{align}
where $\tau$ denotes a relaxation time scale and $X_{ij}$ represents a source term.
For example, we consider the featureless UR of the TC flow and choose the strain rate $C S_{ij}$ with a constant $C$ as a source term $X_{ij}$;~namely, $X_{ij}=CS_{ij}$. 
This model corresponds to the eddy-viscosity model when neglecting the history effect.
For this case, the Reynolds shear stress will yield (see also Refs.~\citep{hd2008,hamba2017,ariki2017})
\begin{align}
    R_{r\theta} = K b_{r\theta} = - \frac{C}{1 + (\tau U_\theta/r)^2} \tau K S_{r\theta}.
    \label{eq:lagrangianhistorymodel}
\end{align}
This model is essentially similar to the model given by Eq.~(\ref{eq:ccarsm}) with $C_3=0$, and thus has the potential to predict the nearly constant mean angular momentum.
However, the model based on the history along the Lagrangian derivative, such as Eq.~(\ref{eq:lagrangianhistorymodel}),
does not form a tensor even if the source $X_{ij}$ is a tensor, because of Eq.~(\ref{eq:lagrangianderivativenottensor}).
Hence, the physical understanding of the resultant models, such as those given by Eq.~(\ref{eq:lagrangianhistorymodel}), will depend on the coordinate system chosen.
We have to employ a covariant time derivative and related time integration to rigorously understand the history effect on tensors in a frame-independent manner.

\citet{ariki2017} discussed the covariance of ARSM in a cylindrical coordinate based on the time integration of the mean strain rate with a relaxation time expressed by the turbulence time scale.
They demonstrated that the integration of the mean strain rate along the upper-convected or Oldroyd derivative yields a covariant linear eddy-viscosity model, in contrast to that along the Lagrangian derivative. 
However, the derived model does not incorporate the correction due to the flow curvature;~that is, the $U_\theta/r$ term in the denominator of Eq.~(\ref{eq:implicitarsm}) does not appear. 
Therefore, we should consider a different mechanism to predict the nearly constant mean angular momentum observed in the TC flows.

\subsubsection{Jaumann derivative}\label{sec:introductionofjaumannderivative}

To construct a covariant ARSM that incorporates convection effects, we employ the Jaumann derivative as an alternative candidate for the covariant time derivative.
Specifically, we employ the Jaumann derivative along the mean velocity, which is hereafter referred to as the mean Jaumann derivative.
The mean Jaumann derivative of a second-rank tensor is defined as follows (see e.g., Refs.~\citep{oldroyd1958,gm1966}; for more general case, see Ref.~\citep{thiffeault2001}):
\begin{align}
    \frac{\mathscr{D} A_{ij}}{\mathscr{D} t} 
    = \frac{\D A_{ij}}{\D t} - W_{i\ell} A_{\ell j} - W_{j\ell} A_{i\ell}.
    \label{eq:meanjaumannderivative}
\end{align}
Note that $W_{ij}$ on the right-hand side is not the absolute vorticity tensor $W^\mathrm{A}_{ij}$ but is the mean vorticity, $W_{ij} = (\nabla_j U_i - \nabla_i U_j)/2$, even in a rotating frame.
The modeled transport equation for the anisotropy tensor given by Eq.~(\ref{eq:reynoldsstresstransportmodel}) can be rewritten via the mean Jaumann derivative as
\begin{align}
    \frac{\mathscr{D} b_{ij}}{\mathscr{D} t} 
    & = - \left(C_\mathrm{S} - 1 + \frac{P^K}{\varepsilon} \right)\frac{\varepsilon}{K} b_{ij} 
    - \left(\frac{4}{3} - C_\mathrm{R1} \right) S_{ij}
    \nonumber \\
    & \hspace{1em}
    - (1 - C_\mathrm{R2}) \left[ S_{i\ell} b_{\ell j} + S_{j\ell} b_{\ell i} \right]_\mathrm{tl}
    - (2 - C_\mathrm{R3}) \left( W^\mathrm{A}_{i \ell} b_{\ell j} + W^\mathrm{A}_{j \ell} b_{\ell i} \right).
    \label{eq:jaumanntransport}
\end{align}
According to previous ARSMs \citep{taulbee1992,wj2000} and our results, setting $C_\mathrm{R2} = 1$ is physically acceptable.
Thus, the transport equation is reduced to
\begin{align}
    \frac{\mathscr{D} b_{ij}}{\mathscr{D} t} 
    = - \left(C_\mathrm{S} - 1 + \frac{P^K}{\varepsilon} \right)\frac{\varepsilon}{K} b_{ij} 
    - \left(\frac{4}{3} - C_\mathrm{R1} \right) S_{ij}
    - (2 - C_\mathrm{R3}) \left( W^\mathrm{A}_{i \ell} b_{\ell j} + W^\mathrm{A}_{j \ell} b_{\ell i} \right).
    \label{eq:jaumanntransportreduced}
\end{align}
If we adopt the mean Oldroyd derivative instead of the Jaumann derivative, the transport equation for the anisotropy tensor yields (see also Refs.~\citep{ariki2015,ariki2017})
\begin{align}
    \frac{\mathfrak{D} b_{ij}}{\mathfrak{D} t} 
    & = - \left(C_\mathrm{S} - 1 + \frac{P^K}{\varepsilon} \right)\frac{\varepsilon}{K} b_{ij} 
    - \left(\frac{4}{3} - C_\mathrm{R1} \right) S_{ij}
    \nonumber \\
    & \hspace{1em}
    - (2 - C_\mathrm{R2}) \left[ S_{i\ell} b_{\ell j} + S_{j\ell} b_{\ell i} \right]_\mathrm{tl}
    - (2 - C_\mathrm{R3}) \left( W^\mathrm{A}_{i \ell} b_{\ell j} + W^\mathrm{A}_{j \ell} b_{\ell i} \right).
    \label{eq:oldroydtransport}
\end{align}
Here, $\mathfrak{D}/\mathfrak{D} t$ denotes the mean Oldroyd derivative;~the mean Oldroyd derivative for a second-rank tensor yields
\begin{align}
    \frac{\mathfrak{D} A_{ij}}{\mathfrak{D} t}
    = \frac{\D A_{ij}}{\D t} - (\nabla_\ell U_i) A_{\ell j} - (\nabla_\ell U_j) A_{i\ell}.
\end{align}
Comparing Eq.~(\ref{eq:oldroydtransport}) with Eq.~(\ref{eq:jaumanntransport}), the coefficient of the third term on the right-hand side yields $2-C_\mathrm{R2}$.
Thus, the contribution of strain rates to the transport equation cannot be neglected as done in Eq.~(\ref{eq:jaumanntransportreduced}), even if we set $C_\mathrm{R2}=1$.
To simplify the Reynolds stress transport while holding the covariance, employing the mean Jaumann derivative is better than doing the mean Oldroyd derivative.

\subsubsection{Application to circular flows}\label{eq:application}

Consider a statistically two-dimensional flow with a circular mean velocity expressed by $U_\theta = U_\theta (r)$ as written in Eq.~(\ref{eq:meanvelocitycondition}).
When $K$, $\varepsilon$, and $g$ are constant along the mean streamline, we can integrate Eq.~(\ref{eq:jaumanntransportreduced}).
The derivation is given in Appendix~\ref{sec:derivationofmodel}.
As a result, the Reynolds shear stress in the inertial frame is expressed as follows:
\begin{align}
    R_{r\theta} & = - \frac{2C_1}{1 + 4(g \tau)^2 (W_{r\theta} + U_\theta/r)^2} 
    \nonumber \\
    & \hspace{2em} \times
    \left[ 1 + \frac{8C_{\mathscr{D}3}}{1 + 4(g \tau)^2 (W_{r\theta} + U_\theta/r)^2} (g \tau)^2 
    \left( W_{r\theta} + \frac{U_\theta}{r} \right) W_{r\theta} \right]
    g \tau K S_{r\theta}.
    \label{eq:jaumannmodelintegrated}
\end{align}
We refer to this model as the time-integrated Jaumann-derivative model (TIJDM).
Comparing Eq.~(\ref{eq:jaumannmodelintegrated}) with the linear eddy-viscosity model expressed by Eqs.~(\ref{eq:eddyviscosity}) and (\ref{eq:lineareddyviscosity}), a correction term $(g\tau)^2 (W_{r\theta} + U_\theta/r)^2$ appears in the denominator.
Note that $W_{r\theta} + U_\theta/r = -S_{r\theta}$; thus, this correction is the same as that using the strain rate discussed in Appendix~\ref{sec:modelparameter}, which is ineffective in predicting the nearly constant mean angular momentum.
Hence, in the TIJDM, the numerator of the correction term in the square brackets $[\cdots]$ is expected to significantly contribute to the mean angular momentum transport.
Note that $(W_{r\theta} + U_\theta/r) W_{r\theta}$ in the square brackets $[\cdots]$ originates from $W^\mathrm{A}_{i \ell} b_{\ell j} + W^\mathrm{A}_{j \ell} b_{\ell i}$ in Eq.~(\ref{eq:jaumanntransportreduced}) and we expect that this term explains the emergence of the nearly constant mean angular momentum.
This result contrasts with the model considering the history along the Lagrangian derivative given by Eq.~(\ref{eq:lagrangianhistorymodel}) because it considers only the strain rate $S_{ij}$ as the source term.

The model given by Eq.~(\ref{eq:jaumannmodelintegrated}) is obtained by fully integrating Eq.~(\ref{eq:jaumanntransportreduced}).
Although the integration is formally correct, its implementation in numerical simulation is difficult in practice.
We can derive the simplified time-local expression by performing a derivative expansion of the integrands \citep{hd2008}.
The derivation is given in Appendix~\ref{sec:derivationofmodel}.
The obtained simplified model for the shear stress in the inertial frame is expressed as follows:
\begin{align}
    R_{r\theta} 
    & =
    - 2C_1 g \tau K S_{r\theta} 
    + 4 C_1 C_{\mathscr{D}3} (g\tau)^3 K \frac{\mathscr{D}}{\mathscr{D} t} (S_{r\ell} W_{\ell \theta} + S_{\theta \ell} W_{\ell r})
    \nonumber \\
    & =
    - 2C_1 g \tau K S_{r\theta} 
    - 16 C_1 C_{\mathscr{D}3} (g\tau)^3 K \left( W_{r\theta} + \frac{U_\theta}{r} \right) W_{r\theta} S_{r\theta}.
    \label{eq:jaumannmodel}
\end{align}
We refer to this model simply as the Jaumann derivative model (JDM). 
Note that $(W_{r\theta} + U_\theta/r) W_{r\theta} S_{r\theta}$, which originates from the Jaumann derivative of $S_{i\ell} W^\mathrm{A}_{\ell j} + S_{j\ell} W^\mathrm{A}_{\ell i}$ (see Appendix~\ref{sec:derivationofmodel} for details), is common to both the JDM and TIJDM.
\citet{hamba2006} succeeded in predicting the nearly zero mean absolute vorticity by employing a similar model to the JDM, although they employed the Lagrangian derivative transformed into a rotating frame instead of the Jaumann derivative.
As discussed in Sec.~\ref{sec:constantangularmomentum}, the zero mean absolute vorticity in a rotating frame corresponds to constant angular momentum in an inertial frame. 

By comparing Eqs.~(\ref{eq:implicitarsm}) and (\ref{eq:ccarsm}), we may have $g = C_\tau f_\nu = C_\nu f_\nu/C_1$ and $C_1 g \tau K = \nu_\mathrm{T}$.
Consequently, the shear stress component of the TIJDM yields
\begin{align}
    R_{r\theta} & = - \frac{2\nu_\mathrm{T}}{1 + 4 C_\tau^2 f_\nu^2 (K/\varepsilon)^2 (W_{r\theta} + U_\theta/r)^2} 
    \nonumber \\
    & \hspace{2em} \times
    \left[ 1 + \frac{2C_\mathrm{J} f_\nu^2}{1 + 4 C_\tau^2 f_\nu^2 (K/\varepsilon)^2 (W_{r\theta} + U_\theta/r)^2}
    \frac{K^2}{\varepsilon^2} \left( W_{r\theta} + \frac{U_\theta}{r} \right) W_{r\theta} \right] S_{r\theta}.
    \label{eq:jaumannshearstressintegrated}
\end{align}
and that for the JDM yields
\begin{align}
    R_{r\theta} = - 2\nu_\mathrm{T} \left[ 1 + 2C_\mathrm{J} f_\nu^2 \frac{K^2}{\varepsilon^2} \left( W_{r\theta} + \frac{U_\theta}{r} \right) W_{r\theta} \right] S_{r\theta},
    \label{eq:jaumannshearstress}
\end{align}
where $C_\mathrm{J} = 4C_{\mathscr{D}3} C_\tau^2$ is a constant parameter.
The TIJDM reduces to the JDM when $C_\tau = 0$ in the denominator of Eq.~(\ref{eq:jaumannshearstressintegrated}).
In addition, the JDM also reduces to the AKN model when $C_\mathrm{J}=0$.

\subsection{Verification of TIJDM and JDM}\label{sec:verificationjaumannmodel}

We verify the performance of the TIJDM and JDM in the TC flow. 
The numerical procedure is the same as that described in Sec.~\ref{sec:numericalsetup}. 
In the convergence calculation, we employ the following regularization to avoid numerical divergence due to negative diffusion:
\begin{gather}
    R_{r\theta} = -2\nu_\mathrm{T} f_\mathrm{I} f_\mathrm{J} S_{r\theta}, \ \ 
    f_\mathrm{J} = 
    \max \left[ 1 + 2C_\mathrm{J} f_\nu^2 f_\mathrm{I} \frac{K^2}{\varepsilon^2} \left( W_{r\theta} + \frac{U_\theta}{r} \right) W_{r\theta}, 0.01 \right], 
    \label{eq:jdm}
\end{gather}
where
\begin{align}
    f_\mathrm{I} = 
    \frac{1}{1 + 4 C_\tau^2 f_\nu^2 (K/\varepsilon)^2 (W_{r\theta} + U_\theta/r)^2},
    \label{eq:jdmcorrection}
\end{align}
for the TIJDM, whereas $f_\mathrm{I} = 1$ for the JDM.
Note that we confirmed that the converged results yielded $f_\mathrm{J} > 0.01$ for all cases performed (figure not shown).
We numerically solved Eqs.~(\ref{eq:ranseq}), (\ref{eq:keq}), and (\ref{eq:epsiloneq}) with Eqs.~(\ref{eq:eddyviscosity}), (\ref{eq:jdm}), and (\ref{eq:jdmcorrection}). 
The TIJDM has two additional parameters $C_\mathrm{J}$ and $C_\tau$, whereas the JDM has only one additional parameter $C_\mathrm{J}$.
Hereafter, we fix $C_\tau$ as the same value as the ccARSM performed in Sec.~\ref{sec:verification}, $C_\tau = 1/3.9$, and change only $C_\mathrm{J}$ to observe the qualitative effects of the Jaumann derivative terms.

\begin{figure}
    \centering
    \begin{minipage}[t]{.48\linewidth}
        \centering
        \includegraphics[width=\linewidth]{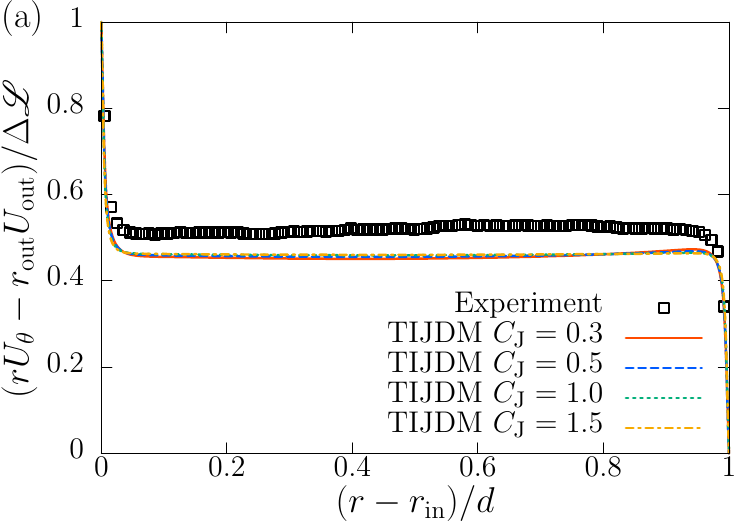}
    \end{minipage}
    \hspace{.01\linewidth}
    \begin{minipage}[t]{.48\linewidth}
        \centering
        \includegraphics[width=\linewidth]{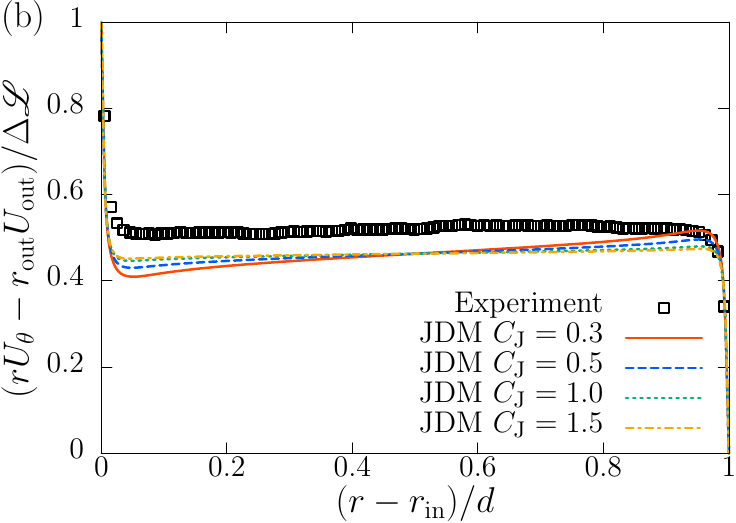}
    \end{minipage}
     \caption{Model parameter $C_\mathrm{J}$ dependence of (a) TIJDM and (b) JDM on the mean angular momentum prediction at $\Rey_\mathrm{in}=8.5\times 10^4$ for $a=-0.33$.}
\label{fig:angularjaumannpara}
\end{figure}

Figure~\ref{fig:angularjaumannpara} depicts the model parameter $C_\mathrm{J}$ dependence of the TIJDM and JDM in predicting the mean angular momentum.
The TIJDM predicts the nearly constant profile regardless of the $C_\mathrm{J}$ value.
For the JDM, the profiles tend to become constant as $C_\mathrm{J}$ increases and are almost saturated at $C_\mathrm{J}=1.0$.
These results suggest that the Jaumann derivative of $S_{i\ell} W^\mathrm{A}_{\ell j} + S_{j\ell} W^\mathrm{A}_{\ell i}$ term, whose shear stress component is expressed by the second part of Eq.~(\ref{eq:jaumannmodel}) in an inertial frame of cylindrical coordinates, is effective in predicting the nearly constant angular momentum.
Furthermore, the success of the JDM indicates that retaining the first time derivative is enough to represent the curvature effects emanating from the history effects in the TC flow.
Hereafter, we fix $C_\mathrm{J}=1.0$.

\begin{figure}
    \centering
    \begin{minipage}[t]{.48\linewidth}
        \centering
        \includegraphics[width=\linewidth]{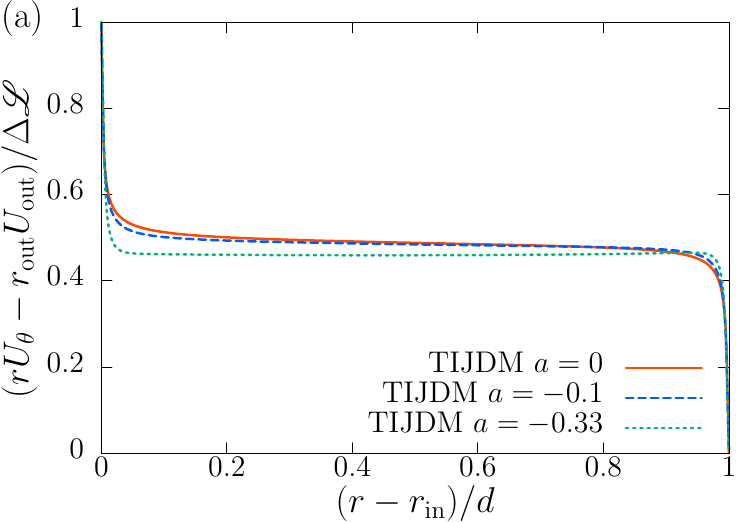}
    \end{minipage}
    \hspace{.01\linewidth}
    \begin{minipage}[t]{.48\linewidth}
        \centering
        \includegraphics[width=\linewidth]{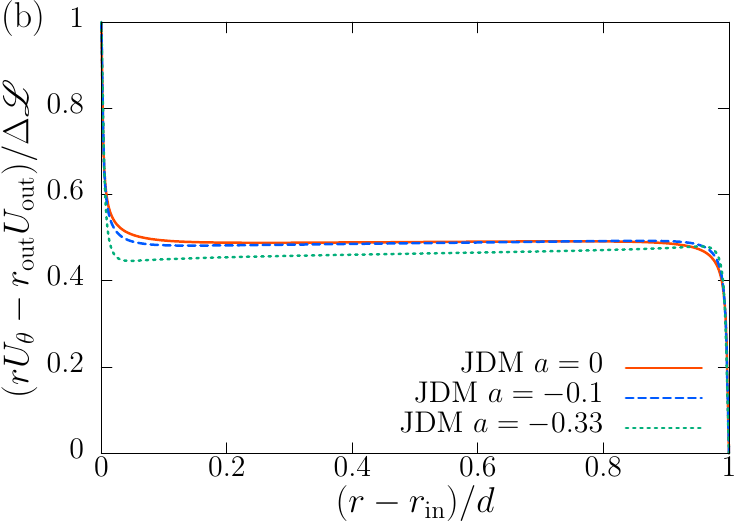}
    \end{minipage}
    \caption{Mean angular momentum profiles of (a) TIJDM and (b) JDM at $\Rey_\mathrm{in}=8.5\times 10^4$ for $a=0$, $-0.1$, and $-0.33$.}
\label{fig:angularjaumann}
\end{figure}

Figure~\ref{fig:angularjaumann} depicts the mean angular momentum profiles for the TIJDM and JDM for several values of $a$.
For both models, the profiles are nearly constant in the bulk region and collapse to approximately $0.5$ for all the cases.
Therefore, we conclude that the Jaumann derivative of $S_{i\ell} W^\mathrm{A}_{\ell j} + S_{j\ell} W^\mathrm{A}_{\ell i}$ term accounts for the emergence of the nearly constant angular momentum in the featureless UR of the TC flows.

\subsection{Interpretation of history effects in JDM}\label{sec:historyeffect}

As we have seen in Sec.~\ref{sec:verificationjaumannmodel}, the JDM, which retains only the first time derivative in the TIJDM, is enough to explain the emergence of the nearly constant angular momentum.
We provide a physical interpretation of the history effects embedded in the JDM.

Consider fluid particles put on a uniform grid.
The grid composed of the fluid particles will be distorted when the velocity field is non-uniform.
This indicates that the coordinates formed by the fluid particles will change due to the convection.
The Oldroyd derivative takes into account the distortion of the coordinates along the fluid path.
In contrast, the Jaumann derivative takes into account corrections arising solely from the generation of the different components of vectors or tensors due to the local vorticity.
For example, let us consider the statistically steady flow in an inertial frame of cylindrical coordinates with the mean velocity $U_\theta(r)$.
In this case, the mean Jaumann derivative of radial and azimuthal components of a steady vector, $A_r(r)$ and $A_\theta(r)$, yields
\begin{align}
    \frac{\mathscr{D} A_r}{\mathscr{D} t} & = - U_\theta \frac{A_\theta}{r} - W_{r\theta} A_\theta, 
    \label{eq:meanjaumannr} \\
    \frac{\mathscr{D} A_\theta}{\mathscr{D} t} & = U_\theta \frac{A_r}{r} - W_{\theta r} A_r. 
    \label{eq:meanjaumannt}
\end{align}
The first terms on the right-hand side of Eqs.~(\ref{eq:meanjaumannr}) and (\ref{eq:meanjaumannt}) represent the effect of convection, whereas the second terms represent the effect of rotation.
The latter terms remain even at the stagnation points $U_\theta = 0$.
The additional terms in the Jaumann derivative compared with the Lagrangian derivative account for the effect of local rotation of the coordinates, which guarantees the covariance.

For the case of TC flows, the velocity fluctuations are anisotropic \cite{ezetaetal2018}.
Therefore, the normal components of the anisotropy tensor, $b_{rr}$ and $b_{\theta \theta}$, are nonzero, and their profiles are different from each other.
We demonstrate that the normal stress difference modifies the Reynolds shear stress via the history effects.
Assume a statistically steady flow.
When we write the source term of the normal stress difference of $b_{ij}$ as $X_{ij}$, which means $X_{\theta \theta} - X_{rr} \neq 0$, the transport equation of $b_{ij}$ along the mean Jaumann derivative with the relaxation time $\tau$ can be written as
\begin{align}
    \frac{\mathscr{D} b_{ij}}{\mathscr{D} t} = - \frac{b_{ij}}{\tau} + X_{ij}.
    \label{eq:jaumannsimplifized}
\end{align}
The model expression that corresponds to the JDM can be written as follows:
\begin{align}
    b_{ij} = \tau X_{ij} - \tau^2 \frac{\mathscr{D} X_{ij}}{\mathscr{D} t}.
\end{align}
Thus, for the TC flows, the shear stress component yields
\begin{align}
    b_{r\theta} = \tau X_{r\theta} - \tau^2 \frac{\mathscr{D} X_{r\theta}}{\mathscr{D} t}
    & = \tau X_{r\theta} - \tau^2 \left[ 
    - U_\theta \frac{X_{\theta \theta} - X_{rr}}{r} 
    - W_{r\theta} X_{\theta \theta}
    + W_{\theta r} X_{rr} \right]
    \nonumber \\
    & = \tau X_{r\theta} - \tau^2 \left[ 
    - U_\theta \frac{X_{\theta \theta} - X_{rr}}{r} 
    - W_{r\theta} ( X_{\theta \theta} - X_{rr} )
    \right].
\end{align}
This result indicates that the source of normal stress difference $X_{\theta \theta} - X_{rr}$ will modify the Reynolds shear stress via the history effects.
In turbulence modeling, $S_{i\ell} W^\mathrm{A}_{\ell j} + S_{j\ell} W^\mathrm{A}_{\ell i}$ is considered as a primitive term representing the source of normal stress difference (see e.g., Refs.~\cite{wj2000,yoshizawabook,taulbee1992}).
Our present model, provided by Eq.~(\ref{eq:jaumannmodel}) or (\ref{eq:jaumannshearstress}), takes into account the time history of $S_{i\ell} W^\mathrm{A}_{\ell j} + S_{j\ell} W^\mathrm{A}_{\ell i}$ through the mean Jaumann derivative, which leads to the success in predicting the nearly constant angular momentum in the featureless UR of the TC turbulence.

The normal stress difference is widely observed in rotating or curved turbulent flows \cite{grundestametal2008,ka2016jfm,brethouwer2022}.
As mentioned before, these flows exhibit nearly zero mean absolute vorticity or constant mean angular momentum states.
\citet{hamba2006} also discussed the significance of the normal stress difference for explaining the zero mean absolute vorticity state in the spanwise rotating channel flows.
They also demonstrated that the temporally nonlocal effect, which is equivalent to the time history effect, provides the interpretation of the model accompanied by the time derivative term, by using the Green's function.
Note that the Green's function used in that study is based on the Lagrangian derivative transformed into a rotating frame, which is covariant only under the Euclidean transformation, but is not generally covariant.
The Jaumann derivative guarantees the general covariance, and thus, it will provide a physical interpretation independent of the coordinate frames assumed.
The analysis based on the covariant time derivative provides a more reliable suggestion that the history effects for the Reynolds stress or the anisotropy tensor are significant for the physical interpretation of the nearly constant mean angular momentum or zero mean absolute vorticity states.

\subsection{Comparison with rotating plane parallel flows}\label{sec:rotatingplanecouette}

The JDM, whose general expression is given by Eq.~(\ref{eq:jaumannmodelgeneral}), is straightforwardly applicable to different geometries and frames owing to the covariance or form invariance of the derivation.
For example, we consider flows between two parallel plates in the Cartesian coordinate system with the parallel mean velocity $\bm{U} = (U_x, U_y, U_z) = (U_x(y), 0, 0)$ where $x$, $y$, and $z$ are the streamwise, wall-normal, and spanwise directions.
In a rotating frame with $\Omega^\mathrm{F}_z = \mathrm{const.}$, the JDM provides the following Reynolds shear stress $R_{xy}$ expression:
\begin{align}
    R_{xy} 
    = - 2 \nu_\mathrm{T,eff} S_{xy},
    \label{eq:jaumannmodelincartesian}
\end{align}
where
\begin{align}
    \nu_\mathrm{T,eff} = C_1 g \tau K \left[ 1 + 8 C_{\mathscr{D}3} (g\tau)^2 W_{xy} W_{xy}^\mathrm{A} \right].
    \label{eq:effectiveviscosityinrpcf}
\end{align}
Under the zero mean absolute vorticity states, $W_{xy} = 2\Omega^\mathrm{F}_z$, this model is equivalent to that proposed by \citet{hamba2006}:
\begin{align}
    R_{xy} = - 2 \nu_\mathrm{T} \left[ 1 + C_{\nu 3} \frac{K^2}{\varepsilon^2} \Omega^\mathrm{F}_z W_{xy}^\mathrm{A} \right] S_{xy},
\end{align}
where $\nu_\mathrm{T}$ is the same as that in the linear eddy-viscosity model given by Eq.~(\ref{eq:eddyviscosity}).
This model succeeded in predicting the zero mean absolute vorticity states.
Furthermore, the models share a similar origin as both consider the time history of the $S_{i\ell} W^\mathrm{A}_{\ell j} + S_{j\ell} W^\mathrm{A}_{\ell i}$ term.
The difference is the form of the time derivative; we consider the Jaumann derivative, whereas \citet{hamba2006} does the Lagrangian derivative.
The Jaumann derivative is preferable for deriving covariant turbulence models.

In the rotating plane Couette flows, the sustainment of zero mean absolute vorticity can be explained in terms of the response to the perturbation of the mean velocity gradient \citep{ka2016jfm}.
The zero mean absolute vorticity indicates that $\d U_x/\d y = 2\Omega^\mathrm{F}_z$.
Thus, when $\Omega^\mathrm{F}_z > 0$, $\d U_x/ \d y > 0$.
Here, we consider a small perturbation from the zero mean absolute vorticity;
\begin{align}
    \frac{\d U_x}{\d y} = 2 \Omega^\mathrm{F}_z \to \frac{\d U_x}{\d y} = 2\Omega^\mathrm{F}_z + \Delta
\end{align}
where $\Delta$ is the small perturbation.
Note that when $\Omega^\mathrm{F}_z > 0$, $\Delta > 0$ increases the velocity gradient, whereas $\Delta < 0$ decreases it.
According to Eq.~(\ref{eq:jaumannmodelincartesian}), the modification of $\nu_\mathrm{T,eff}$ due to $\Delta$ is evaluated as
\begin{align}
    \delta \nu_\mathrm{T,eff} \propto W_{xy} W^\mathrm{A}_{xy} 
    = \frac{1}{2} \Omega^\mathrm{F}_z \Delta,
\end{align}
where $O(\Delta^2)$ term is neglected.
If $\Delta > 0$ and $\Omega^\mathrm{F}_z > 0$, $\delta \nu_\mathrm{T,eff} > 0$, which decreases the mean velocity gradient due to the increase in effective viscosity.
Thus, the effective viscosity is modified as it cancels the perturbation $\Delta > 0$.
On the other hand, if $\Delta < 0$ and $\Omega^\mathrm{F}_z > 0$, $\delta \nu_\mathrm{T,eff} < 0$, which increases the mean velocity gradient, and this effect also cancels the perturbation $\Delta < 0$.
This cancellation does not change even if $\Omega^\mathrm{F}_z < 0$.
Therefore, for the JDM, the zero mean absolute vorticity is stable against the perturbation $\Delta$.
This scenario is essentially the same as that proposed by \citet{ka2016jfm}, although they demonstrated this in terms of the Reynolds stress transport.
Therefore, the JDM can explain the sustainment of the zero mean absolute vorticity in the rotating plane Couette flows.

Note that the same scenario is applicable to the constant mean angular momentum in the TC flows.
For the TC flows in the featureless ultimate regime, we have
\begin{align}
    \nu_\mathrm{T,eff} 
    = C_1 g \tau K \left[ 1 + 8 C_{\mathscr{D}3} (g\tau)^2 \left( W_{r\theta} + \frac{U_\theta}{r} \right) W_{r\theta} \right],
\end{align}
from Eq.~(\ref{eq:jaumannmodel}).
We consider the small perturbation from the constant mean angular momentum in the following form
\begin{align}
    W_{r\theta} = - \frac{1}{2r} \frac{\d}{\d r} (r U_\theta) = 0
    \to
    W_{r\theta} = \Delta.
\end{align}
Here, $\Delta > 0$ indicates that $U_\theta$ has a steeper slope than the constant mean angular momentum state, whereas $\Delta < 0$ indicates that $U_\theta$ has a shallower slope.
The modification of $\nu_\mathrm{T.eff}$ due to $\Delta$ yields
\begin{align}
    \delta \nu_\mathrm{T,eff} \propto \left( W_{r\theta} + \frac{U_\theta}{r} \right) W_{r\theta}
    = \frac{U_\theta}{r} \Delta.
\end{align}
For the co-rotating case, $U_\theta > 0$.
Therefore, if $\Delta > 0$, $\delta \nu_\mathrm{T,eff} > 0$, which makes the slope of $U_\theta$ shallow.
On the other hand, if $\Delta < 0$, $\delta \nu_\mathrm{T,eff} < 0$, which makes the slope of $U_\theta$ steep.
Therefore, for the JDM, the constant mean angular momentum is stable against the perturbation $\Delta$.
Then, we can conclude that there exists a common mechanism of the sustainment for the constant mean angular momentum profile in Taylor--Couette flows and the zero mean absolute vorticity in rotating plane Couette flows.

\section{Conclusions}\label{sec:conclusions}

The mechanism of the emergence of the nearly constant angular momentum state in the featureless ultimate regime (UR) of the Taylor--Couette (TC) turbulence was discussed in terms of history effects.
We employed the Reynolds-averaged Navier--Stokes (RANS) models to find an essential contribution to the angular momentum under the statistical average.

The experimental results showed universal profiles of a nearly constant mean angular momentum in the bulk region, mainly for the co-rotating cases, which is consistent with the numerical results of \citet{brauckmannetal2016}.
For the RANS simulation, the conventional linear eddy-viscosity model did not predict the nearly constant mean angular momentum for the co-rotating case, although it worked fairly well for the case in which only the inner cylinder rotated.
In contrast, the ARSMs accompanied by the convection effects in the Reynolds stress transport predicts the mean angular momentum profiles well for both cases.

The failure of the linear eddy-viscosity model was discussed in terms of the scaling of the dissipation rate.
The spatial profiles of the dissipation rate for the linear eddy-viscosity model in the TC flows were consistent with those expected from the logarithmic region near the wall for plane shear flows.
This failure results from the lack of convection effects in the Reynolds stress transport, according to the results of the ARSMs tested.

To incorporate the convection effects with the Reynolds stress in a coordinate-independent form, we employed the Jaumann derivative as a covariant time derivative \citep{oldroyd1958,gm1966,thiffeault2001,frewer2009}.
The small parameter expansion based on the nearly constant angular momentum led to the ARSM employing time-integral terms.
The resultant model was named the time-integrated Jaumann derivative model (TIJDM).
To simplify the TIJDM, we performed a derivative expansion on the integrand and obtained the model by retaining the terms up to the first time derivative, named the JDM.

Both the TIJDM and JDM succeeded in predicting nearly constant angular momentum profiles for co-rotating turbulent TC flows, including the case in which only the inner cylinder rotates.
Thus, the essence of predicting the nearly constant mean angular momentum observed in the featureless UR of the TC turbulence is embedded in the first time derivative terms in the simplified model, JDM.
We demonstrated that the success of the JDM originates from the time history effects of the normal stress difference.
In addition, we confirmed that the JDM explains the sustainment of both the zero mean absolute vorticity in the rotating plane shear flows and the constant mean angular momentum in TC flows in the same way.
This suggests that their mechanism is common.
Turbulent shear flows are usually anisotropic, including rotating or curved turbulent flows \cite{grundestametal2008,ka2016jfm,brethouwer2022} where the nearly zero mean absolute vorticity or constant mean angular momentum states emerge.
This study suggests the significance of the history effects of the Reynolds stress or the anisotropy tensor for understanding curved or rotating turbulent flows in terms of the statistical analysis of turbulent flows.

This study focused on the featureless UR regime in TC turbulence.
\citet{grossmannetal2016} suggested that the Taylor roll structures are significant for the optimal angular velocity transport.
Thus, the featureless UR is not desirable for optimal angular velocity transport.
Persistent existence of the Taylor rolls changes the history or transport path of the Reynolds stress and alters the mean angular momentum profiles, which suggests a change in the efficiency of the angular velocity transport.
The analysis of the history effects of the Reynolds stress due to the Taylor rolls is our future work.

\begin{acknowledgments}
The authors acknowledge Prof.~Hiromichi Kobayashi at Keio University for valuable discussions and Mr.~Homare Okuyama for his experimental support.
K. I. was supported by Grant-in-Aid for JSPS Fellows Grant No.~JP22KJ2660 and Grant-in-Aid from the Harris Science Research Institute of Doshisha University. 
Y. H. was supported by JSPS KAKENHI (C) Grant No.~JP22K03917.
\end{acknowledgments}

\appendix

\section{Covariant form of spatial gradient in cylindrical coordinate}\label{sec:covariantderivative}

We write the velocity field in cylindrical coordinates as follows:
\begin{align}
    \bm{u} = u_r \bm{e}_r + u_\theta \bm{e}_\theta + u_z \bm{e}_z,
\end{align}
where $\bm{e}_r$, $\bm{e}_\theta$, and $\bm{e}_z$ are unit vectors in the radial, azimuthal, and axial directions, respectively.
We only consider an inertial frame in this appendix.
We can write the unit vectors in Cartesian coordinates as
\begin{align}
    \bm{e}_r = 
    \begin{bmatrix}
    \cos \theta \\ \sin \theta \\ 0
    \end{bmatrix}
    ,\ \ 
    \bm{e}_\theta = 
    \begin{bmatrix}
    -\sin \theta \\ \cos \theta \\ 0
    \end{bmatrix}
    ,\ \ 
    \bm{e}_z = 
    \begin{bmatrix}
    0 \\ 0 \\ 1
    \end{bmatrix}.
\end{align}
The azimuthal derivatives of $\bm{e}_r$ and $\bm{e}_\theta$ yield nonzero; namely,
\begin{gather}
    \frac{\partial}{\partial \theta} \bm{e}_r = \bm{e}_\theta, \ \ 
    \frac{\partial}{\partial \theta} \bm{e}_\theta = -\bm{e}_r.
\end{gather}
Therefore, we have
\begin{subequations}
\begin{align}
    \frac{\partial}{\partial r} \bm{u} & = \frac{\partial u_r}{\partial r} \bm{e}_r + \frac{\partial u_\theta}{\partial r} \bm{e}_\theta + \frac{\partial u_z}{\partial r}, \\
    \frac{\partial}{\partial \theta} \bm{u} & = \frac{\partial u_r}{\partial \theta} \bm{e}_r + \frac{\partial u_\theta}{\partial \theta} \bm{e}_\theta
    + u_r \bm{e}_\theta - u_\theta \bm{e}_r
    + \frac{\partial u_z}{\partial \theta}.
\end{align}
\end{subequations}
Thus, the covariant form of the velocity gradient in cylindrical coordinates should be
\begin{align}
    \bm{\nabla} \bm{u} =
    \begin{bmatrix}
    \nabla_r u_r & \nabla_\theta u_r & \nabla_z u_r \\
    \nabla_r u_\theta & \nabla_\theta u_\theta & \nabla_z u_\theta \\
    \nabla_r u_z & \nabla_\theta u_z & \nabla_z u_z
    \end{bmatrix}
    = 
    \begin{bmatrix}
    \displaystyle \frac{\partial u_r}{\partial r} & 
    \displaystyle \frac{1}{r} \frac{\partial u_r}{\partial \theta} - \frac{u_\theta}{r} &
    \displaystyle \frac{\partial u_r}{\partial z} \\
    \displaystyle \frac{\partial u_\theta}{\partial r} &
    \displaystyle \frac{1}{r} \frac{\partial u_\theta}{\partial \theta} + \frac{u_r}{r} &
    \displaystyle \frac{\partial u_\theta}{\partial z} \\
    \displaystyle \frac{\partial u_z}{\partial r} &
    \displaystyle \frac{1}{r} \frac{\partial u_z}{\partial \theta} &
    \displaystyle \frac{\partial u_z}{\partial z} 
    \end{bmatrix}.
    \label{eq:covariantvelocitygradient2}
\end{align}
Then, we obtaine Eq.~(\ref{eq:covariantvelocitygradient}).
This can be written using the suffix notation as
\begin{align}
    \nabla_j u_i = \frac{\partial u_i}{\partial x_j} + C_{i\ell,j} u_\ell,
\end{align}
where $(u_1,u_2,u_3) = (u_r, u_\theta, u_z)$ and
\begin{gather}
    \frac{\partial}{\partial x_1} = \frac{\partial}{\partial r}, \ \ 
    \frac{\partial}{\partial x_2} = \frac{1}{r} \frac{\partial}{\partial \theta}, \ \ 
    \frac{\partial}{\partial x_3} = \frac{\partial}{\partial z}, 
    \nonumber \\
    C_{i\ell,j} = \bm{e}_i \cdot \frac{\partial \bm{e}_\ell}{\partial x_j}
    =
    \begin{cases}
    C_{21,2} = - C_{12,2} = \dfrac{1}{r} & \\
    0 & \text{(otherwise)}
    \end{cases},
    \label{eq:christoffelsymbolforcylindricalcoordinates}
\end{gather}
with $(\bm{e}_1,\bm{e}_2,\bm{e}_3) = (\bm{e}_r,\bm{e}_\theta,\bm{e}_z)$. 
$C_{i\ell,j}$ corresponds to the Christoffel symbol.
In addition, the covariant derivative of a second-rank tensor $A_{ij}$ in this formulation yields
\begin{align}
    \nabla_\ell A_{ij} = \frac{\partial A_{ij}}{\partial x_\ell} + C_{im,\ell} A_{mj} + C_{jm,\ell} A_{im}.
\end{align}
Even if we adopt this formulation, we can derive the covariant form of equations using a local orthogonal transformation.
The velocity gradient expressed in Eq.~(\ref{eq:covariantvelocitygradient2}) satisfies the rule of orthogonal transformation; namely, the transformation of the velocity gradient between the Cartesian and cylindrical coordinates yields
\begin{align}
    \begin{bmatrix}
    \dfrac{\partial u_x}{\partial x} & \dfrac{\partial u_x}{\partial y} & \dfrac{\partial u_x}{\partial z} \\
    \dfrac{\partial u_y}{\partial x} & \dfrac{\partial u_y}{\partial y} & \dfrac{\partial u_y}{\partial z} \\
    \dfrac{\partial u_z}{\partial x} & \dfrac{\partial u_z}{\partial y} & \dfrac{\partial u_z}{\partial z}
    \end{bmatrix}
    =
    \mathsf{O}^\mathrm{t}
    \begin{bmatrix}
    \nabla_r u_r & \nabla_\theta u_r & \nabla_z u_r \\
    \nabla_r u_\theta & \nabla_\theta u_\theta & \nabla_z u_\theta \\
    \nabla_r u_z & \nabla_\theta u_z & \nabla_z u_z
    \end{bmatrix}
    \mathsf{O}, 
\end{align}
where $\mathsf{O}^\mathrm{t}$ denotes the transpose of $\mathsf{O}$, and $\mathsf{O}$ is a local orthogonal matrix of transformation defined by
\begin{align}
    \mathsf{O}^\mathrm{t} =
    \begin{bmatrix}
    \dfrac{\partial x}{\partial r} & \dfrac{1}{r} \dfrac{\partial x}{\partial \theta} & \dfrac{\partial x}{\partial z} \\
    \dfrac{\partial y}{\partial r} & \dfrac{1}{r} \dfrac{\partial y}{\partial \theta} & \dfrac{\partial y}{\partial z} \\
    \dfrac{\partial z}{\partial r} & \dfrac{1}{r} \dfrac{\partial z}{\partial \theta} & \dfrac{\partial z}{\partial z}
    \end{bmatrix}
    =
    \begin{bmatrix}
    \cos \theta & - \sin \theta & 0 \\
    \sin \theta & \cos \theta & 0 \\
    0 & 0 & 1
    \end{bmatrix}.
\end{align}
Therefore, the velocity gradient expressed in Eq.~(\ref{eq:covariantvelocitygradient2}) transforms as a tensor under local orthogonal transformation (for details, see e.g., Refs.~\cite{thiffeault2001,frewer2009,ariki2015}).
This calculation rule of a derivative is used for any vector and tensor quantities in cylindrical coordinates. 
For example, the gradients of the second-rank tensor $A_{ij}$ in cylindrical coordinates can be calculated in this formulation as follows:
\begin{subequations}
\begin{align}
    \nabla_\theta A_{rr} & 
    = \frac{1}{r} \frac{\partial A_{rr}}{\partial \theta} 
    + C_{12,2} A_{\theta r} + C_{12,2} A_{r\theta}
    = \frac{1}{r} \frac{\partial A_{rr}}{\partial \theta}
    - \frac{A_{\theta r}}{r} - \frac{A_{r \theta}}{r}, \\
    \nabla_\theta A_{\theta\theta} & 
    = \frac{1}{r} \frac{\partial A_{\theta\theta}}{\partial \theta} 
    + C_{21,2} A_{r\theta} + C_{21,2} A_{\theta r}
    = \frac{1}{r} \frac{\partial A_{\theta\theta}}{\partial \theta} 
    + \frac{A_{r\theta}}{r} + \frac{A_{\theta r}}{r}, \\
    \nabla_\theta A_{r\theta} & 
    = \frac{1}{r} \frac{\partial A_{r\theta}}{\partial \theta} 
    + C_{12,2} A_{\theta \theta} + C_{21,2} A_{rr}
    = \frac{1}{r} \frac{\partial A_{r\theta}}{\partial \theta} 
    - \frac{A_{\theta\theta}}{r} + \frac{A_{rr}}{r}.
\end{align}
\end{subequations}
$\nabla_i u_i$ yields the trace of Eq.~(\ref{eq:covariantvelocitygradient2}). 
Because the metric tensor is the identity operator in this formulation, the Laplacian of the velocity fields $\nabla_j \nabla_j u_i$
can be written as
\begin{align}
    \nabla_j (\nabla_j u_i)
    & = \frac{\partial}{\partial x_j} (\nabla_j u_i) + C_{j\ell,j} \nabla_\ell u_i + C_{i\ell,j} \nabla_j u_\ell
    \nonumber \\
    & = \frac{\partial}{\partial x_j} \left( \frac{\partial u_i}{\partial x_j} + C_{i\ell, j} u_\ell \right) 
    + C_{j\ell, j} \left( \frac{\partial u_i}{\partial x_\ell} + C_{im,\ell} u_m \right) 
    + C_{i\ell, j} \left( \frac{\partial u_\ell}{\partial x_j} + C_{\ell m,j} u_m \right).
\end{align}
Thus, we have
\begin{align}
    \nabla_j \nabla_j 
    \begin{bmatrix}
    u_r \\ u_\theta\\ u_z
    \end{bmatrix}
    =
    \begin{bmatrix}
    \left[ \dfrac{1}{r} \dfrac{\partial}{\partial r} \left( r \dfrac{\partial}{\partial r} \right) + \dfrac{1}{r^2} \dfrac{\partial^2}{\partial \theta^2} + \dfrac{\partial^2}{\partial z^2} \right] u_r
    - \dfrac{2}{r^2} \dfrac{\partial u_\theta}{\partial \theta} - \dfrac{u_r}{r^2} \\
    \left[ \dfrac{1}{r} \dfrac{\partial}{\partial r} \left( r \dfrac{\partial}{\partial r} \right) + \dfrac{1}{r^2} \dfrac{\partial^2}{\partial \theta^2} + \dfrac{\partial^2}{\partial z^2} \right] u_\theta
    + \dfrac{2}{r^2} \dfrac{\partial u_r}{\partial \theta} - \dfrac{u_\theta}{r^2} \\
    \left[ \dfrac{1}{r} \dfrac{\partial}{\partial r} \left( r \dfrac{\partial}{\partial r} \right) + \dfrac{1}{r^2} \dfrac{\partial^2}{\partial \theta^2} + \dfrac{\partial^2}{\partial z^2} \right] u_z
    \end{bmatrix},
\end{align}
in cylindrical coordinates. 
The covariant derivative of scalars, e.g., pressure, yields the conventional partial derivative.
In cylindrical coordinates, the pressure gradient can be written as follows:
\begin{align}
    (\nabla_r p, \nabla_\theta p, \nabla_z p)
    =
    \left(
    \dfrac{\partial p}{\partial r} ,
    \dfrac{1}{r} \dfrac{\partial p}{\partial \theta} ,
    \dfrac{\partial p}{\partial z}
    \right).
\end{align}

\section{Reynolds number dependence of angular momentum}\label{sec:reynoldsnumberdependence}

\begin{figure}[t]
    \centering
    \begin{minipage}[t]{.48\linewidth}
        \centering
        \includegraphics[width=\linewidth]{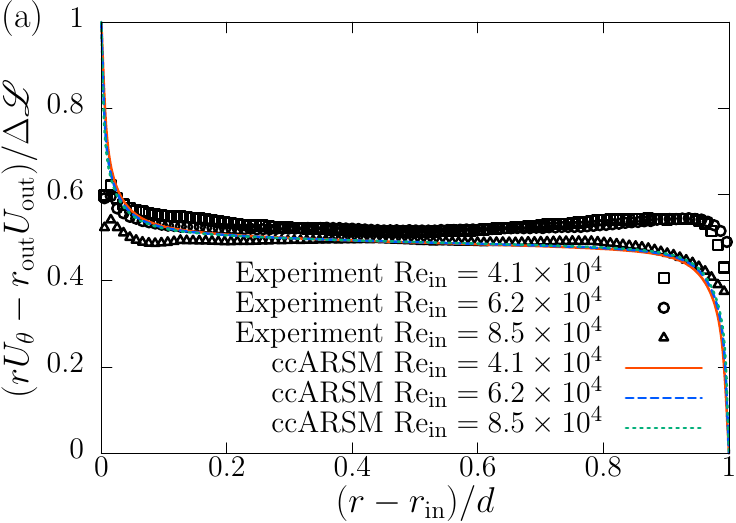}
    \end{minipage}
    \hspace{.01\linewidth}
    \begin{minipage}[t]{.48\linewidth}
        \centering
        \includegraphics[width=\linewidth]{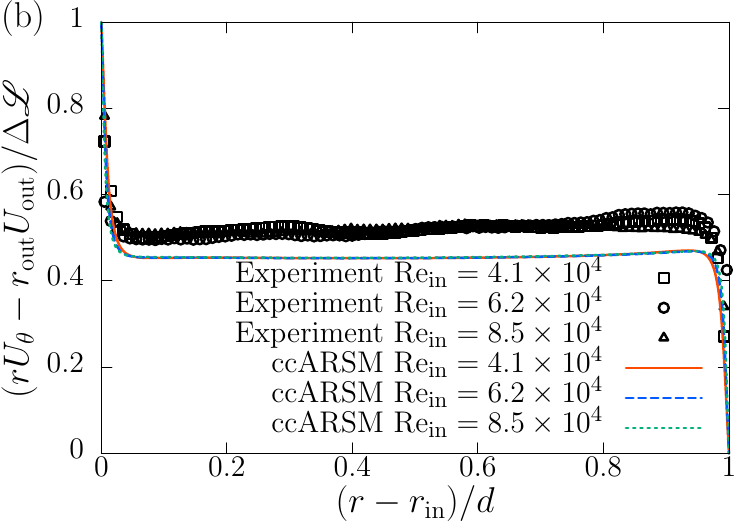}
    \end{minipage}
    \caption{Mean angular momentum profiles of experiments and ccARSM at $\Rey_\mathrm{in}=4.1 \times 10^4$, $6.2\times 10^4$, and $8.5\times 10^4$ for (a) $a=0$ and (b) $a=-0.33$.}
    \label{fig:reynoldsnumberdependence}
\end{figure}

Figure~\ref{fig:reynoldsnumberdependence} shows the mean angular momentum of the experiments and ccARSM at $\Rey_\mathrm{in} = 4.1\times 10^4$, $6.2 \times 10^4$, and $8.5\times10^4$ for $a=0$ and $-0.33$. 
The experimental results are almost constant in the bulk region independently of the Reynolds number for both $a=0$ and $a=-0.33$. 
For the ccARSM, the profile is independent of the Reynolds number, except for a small difference in the vicinity of the wall. 
The RANS model is intrinsically independent of the Reynolds number, based on its concept of construction for predicting high-Reynolds-number turbulent flows. 
Because the bulk region is almost independent of the Reynolds number, the high-Reynolds-number TC turbulence is consistent with the concept of RANS simulation.

\section{Effect of strain rate for predicting constant angular momentum}\label{sec:modelparameter}

To verify the effect of the strain rate in the ARSM given by Eq.~(\ref{eq:implicitarsm}), we consider the following model, similar to Eq.~(\ref{eq:ccarsm}):
\begin{align}
    R_{r\theta} = - \frac{2C_1}{1 + 4 \tau_\mathrm{T}^2 C_2^2 S_{r\theta}^2/3} \tau_\mathrm{T} K S_{r\theta}, \
    \tau_\mathrm{T} = C_\tau f_\nu \frac{K}{\varepsilon}.
    \label{eq:wowr}
\end{align}
We refer to the model given by Eq.~(\ref{eq:wowr}) as the ARSM-S.
We also set $C_2=1$ and $C_1=C_\nu/C_\tau$ with $C_\nu$ such that $C_2 = 0$ leads to the AKN model.
In contrast to Eq.~(\ref{eq:implicitarsm}), the coefficient of the strain rate in the denominator is positive in Eq.~(\ref{eq:wowr}). 
Several studies have employed this type of modeling (see e.g., Refs.~\citep{szl1993,yoshizawaetal2006}).

\begin{figure}
    \centering
    \includegraphics[width=0.5\linewidth]{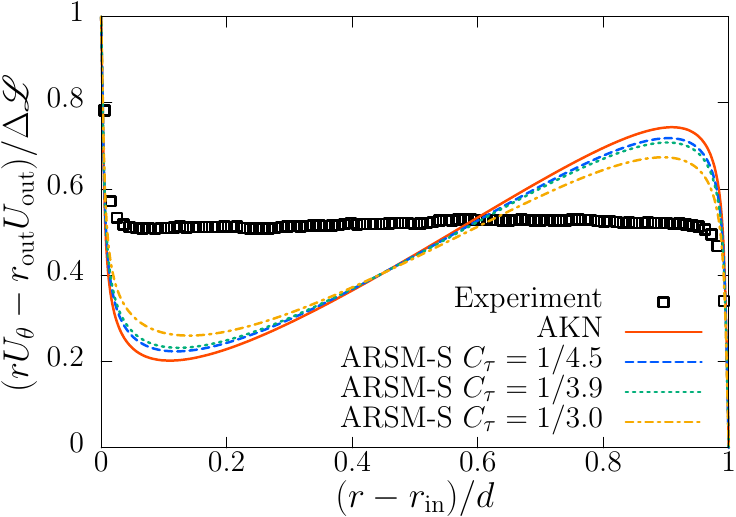}
    \caption{Angular momentum profiles of the AKN (linear eddy-viscosity) model and ARSM-S provided by Eq.~(\ref{eq:wowr}) at $\Rey_\mathrm{in}=8.5\times 10^4$ for $a=-0.33$.}
    \label{fig:angularwowr}
\end{figure}

Figure~\ref{fig:angularwowr} depicts the mean angular momentum predicted by the AKN (linear eddy-viscosity) model and ARSM-S at $\Rey_\mathrm{in}=8.5\times 10^4$ for $a=-0.33$.
As observed in Fig.~\ref{fig:angularwowr}, the ARSM-S is ineffective in predicting a constant mean angular momentum.
Note that for $C_\tau > 1/3$, we cannot numerically obtain the relevant mean velocity or angular momentum profiles.
Consequently, the strain rate in the denominator of the ARSM is not essential for modeling the curvature effects.

\section{Derivation of Jauman derivative models}\label{sec:derivationofmodel}

\subsection{Simple perturbation}\label{sec:perturbation}

To integrate Eq.~(\ref{eq:jaumanntransportreduced}), we perform a simple perturbation analysis based on the following normalization:
\begin{align}
    \widehat{S}_{ij} = \frac{K}{\varepsilon} S_{ij}, \ 
    \widehat{W}^\mathrm{A}_{ij} = \frac{W^\mathrm{A}_{ij}}{\sqrt{W^\mathrm{A}_{\ell m} W^\mathrm{A}_{\ell m}}}, \ 
    \lambda = \frac{K}{\varepsilon} \sqrt{W^\mathrm{A}_{\ell m} W^\mathrm{A}_{\ell m}}, \ 
    \frac{\mathscr{D}}{\mathscr{D} \widehat{t}} = \frac{K}{\varepsilon} \frac{\mathscr{D}}{\mathscr{D} t},
\end{align}
which leads to the following transport equation:
\begin{align}
    g^{-1} b_{ij} + \frac{\mathscr{D} b_{ij}}{\mathscr{D} \widehat{t}} 
    = - 2C_1 \widehat{S}_{ij}
    - C_{\mathscr{D}3} \lambda \left( \widehat{W}^\mathrm{A}_{i \ell} b_{\ell j} + \widehat{W}^\mathrm{A}_{j \ell} b_{\ell i} \right),
    \label{eq:jaumanntransportnormalised}
\end{align}
where $g$ and $C_1$ are defined in Eq.~(\ref{eq:redefinedconstants}), and $C_{\mathscr{D}3} = 2-C_\mathrm{R3} = 1+C_3$.
For a nearly constant mean angular momentum state in an inertial frame or zero mean absolute vorticity state in a rotating frame, $\sqrt{W^\mathrm{A}_{\ell m} W^\mathrm{A}_{\ell m}}$ is very small; thus, $\lambda$ is a small parameter, whereas $\widehat{W}^\mathrm{A}_{ij}$ can be finite.
To incorporate the history effect, we expand the anisotropy tensor as follows:
\begin{gather}
    b_{ij} = b_{ij}^{(0)} + \lambda b_{ij}^{(1)} + \cdots.
    \label{eq:parameterexpansion}
\end{gather}
Up to $O(\lambda)$, we have
\begin{subequations}
\begin{align}
    g^{-1} b_{ij}^{(0)} + \frac{\mathscr{D} b_{ij}^{(0)}}{\mathscr{D} \widehat{t}} 
    & = - 2C_1 \widehat{S}_{ij}, \\
    g^{-1} b_{ij}^{(1)} + \frac{\mathscr{D} b_{ij}^{(1)}}{\mathscr{D} \widehat{t}} 
    & = - C_{\mathscr{D}3} \left( \widehat{W}^\mathrm{A}_{i \ell} b^{(0)}_{\ell j} + \widehat{W}^\mathrm{A}_{j \ell} b^{(0)}_{\ell i} \right).
\end{align}
\end{subequations}
These equations can be formally solved as follows \citep{gm1966,hd2008,hamba2017,ariki2017}:
\begin{subequations}
\begin{align}
    b_{ij}^{(0)} & = - 2C_1 \int^{\widehat{t}} \mathscr{D} \widehat{t}' 
    \exp \left[ -\int^{\widehat{t}}_{\widehat{t}'} \mathscr{D} \widehat{t}'' g^{-1}|_{\widehat{t}''} \right] 
    \widehat{S}_{ij} |_{\widehat{t}'},
    \label{eq:jaumannintegrate1st} \\
    b_{ij}^{(1)} & = - C_{\mathscr{D}3} \int^{\widehat{t}} \mathscr{D} \widehat{t}' 
    \exp \left[ -\int^{\widehat{t}}_{\widehat{t}'} \mathscr{D} \widehat{t}'' g^{-1}|_{\widehat{t}''} \right]  
    \left. \left( \widehat{W}^\mathrm{A}_{i \ell} b^{(0)}_{\ell j} + \widehat{W}^\mathrm{A}_{j \ell} b^{(0)}_{\ell i} \right) \right|_{\widehat{t}'},
    \label{eq:jaumannintegrate2nd}
\end{align}
\end{subequations}
where $\int \mathscr{D}t$ denotes the time integration along the mean Jaumann transport
and $\cdot |_t$ denotes the time label along the transport path:~the path of tensor field along the mean Jaumann derivative.
Although the time-integral expression is formally correct, its implementation in numerical simulation is difficult in practice.
We can derive the simplified time-local expression by performing a derivative expansion of the integrands \citep{hd2008}:
\begin{align}
    A_{ij}|_{\widehat{t}'} 
    = A_{ij}|_{\widehat{t}} 
    - \left. \frac{\mathscr{D} A_{ij}}{\mathscr{D} \widehat{t}} \right|_{\widehat{t}} (\widehat{t}-\widehat{t}') 
    + \frac{1}{2} \left. \frac{\mathscr{D}^2 A_{ij}}{\mathscr{D} \widehat{t}^2} \right|_{\widehat{t}} (\widehat{t}-\widehat{t}')^2
    + \cdots.
    \label{eq:derivativeexpansion}
\end{align}
Consequently, up to $O(\lambda)$ and the first time derivative, we obtain the following algebraic expression for the anisotropy tensor:
\begin{align}
    b_{ij} & = 
    - 2C_1 h_{(0)} \widehat{S}_{ij} 
    + 2C_1 h_{(1)} \frac{\mathscr{D} \widehat{S}_{ij}}{\mathscr{D} \widehat{t}} 
    - 2C_1 C_{\mathscr{D}3} h_{(0)}^2 \lambda \left(\widehat{S}_{i\ell} \widehat{W}^\mathrm{A}_{\ell j} + \widehat{S}_{j\ell} \widehat{W}^\mathrm{A}_{\ell i} \right)
    \nonumber \\
    & \hspace{1em}
    + 2C_1 C_{\mathscr{D}3} h_{(0)} h_{(1)} \lambda \left(\frac{\mathscr{D} \widehat{S}_{i\ell}}{\mathscr{D} \widehat{t}} \widehat{W}^\mathrm{A}_{\ell j} + \frac{\mathscr{D} \widehat{S}_{j\ell}}{\mathscr{D} \widehat{t}} \widehat{W}^\mathrm{A}_{\ell i} \right)
    \nonumber \\
    & \hspace{1em}
    + 2C_1 C_{\mathscr{D}3} h_{(1)} \lambda \frac{\mathscr{D}}{\mathscr{D} \widehat{t}}
    \left[ h_{(0)} \left( \widehat{S}_{i\ell} \widehat{W}^\mathrm{A}_{\ell j} + \widehat{S}_{j\ell} \widehat{W}^\mathrm{A}_{\ell i} \right) \right]
    + \cdots,
    \label{eq:jaumannexpandedmodel}
\end{align}
where 
\begin{align}
    h_{(n)} = \frac{1}{n!} \int^{\widehat{t}} \mathscr{D} \widehat{t}' 
    \exp \left[ -\int^{\widehat{t}}_{\widehat{t}'} \mathscr{D} \widehat{t}'' g^{-1}|_{\widehat{t}''} \right] (\widehat{t}-\widehat{t}')^n.
\end{align}
The expression given by Eq.~(\ref{eq:jaumannexpandedmodel}) can also be derived via the iterative expansion of Eq.~(\ref{eq:jaumanntransport})~\citep{hamba2006}.
For unsteady turbulent flows, $h_{(n)}$ depends on time according to the histories of $K/\varepsilon$ and $g$.

\subsection{Application to circular flows}\label{eq:application}

Consider a statistically two-dimensional flow with a circular mean velocity expressed by $U_\theta = U_\theta (r)$ as written in Eq.~(\ref{eq:meanvelocitycondition}).
When $K$, $\varepsilon$, and $g$ are constant along the mean streamline, we have $h_{(n)} = g^{n+1}$.
Under these assumptions, we can analytically integrate Eqs.~(\ref{eq:jaumannintegrate1st}) and (\ref{eq:jaumannintegrate2nd}) (see e.g., Refs.~\citep{gm1966,hd2008,hamba2017,ariki2019}).
Alternatively, we can obtain the integrated form by performing the derivative expansion given by Eq.~(\ref{eq:derivativeexpansion}) and renormalizing the expanded results via $1-x+x^2-x^3+ \cdots = 1/(1+x)$.
Consequently, the shear stress components yield
\begin{align}
    b^{(0)}_{r\theta} & 
    = - \frac{2C_1}{1 + 4(g\tau)^2 (W_{r\theta} + U_\theta/r)^2} g \widehat{S}_{r\theta}, 
    \label{eq:jaumannintegrated1st} \\
    b^{(1)}_{r\theta} & 
    = - \frac{2C_1}{1 + 4(g\tau)^2 (W_{r\theta} + U_\theta/r)^2} g \widehat{S}_{r\theta} 
    \frac{8C_{\mathscr{D}3}}{1 + 4(g\tau)^2 (W_{r\theta} + U_\theta/r)^2} g^2 \tau \left( W_{r\theta} + \frac{U_\theta}{r} \right) \widehat{W}_{r\theta},
    \label{eq:jaumannintegrated2nd}
\end{align}
where $\tau = K/\varepsilon$.
The difference between the models given by Eqs.~(\ref{eq:lagrangianhistorymodel}) and (\ref{eq:jaumannintegrated1st}) is only the time derivative employed in their derivation;~the former employs the Lagrangian derivative and the latter employs the Jaumann derivative, while the source term is the strain rate for both models.
Note that the correction to the conventional eddy-viscosity model that appears in the denominator yields $W_{r\theta} + U_\theta/r = -S_{r\theta}$.
This correction is the same as that using the strain rate discussed in Appendix~\ref{sec:modelparameter}, and thus it is ineffective in predicting the nearly constant mean angular momentum.
Hence, we have to consider the next order, $O(\lambda)$, term given by Eq.~(\ref{eq:jaumannintegrated2nd}) to explain the emergence of the nearly constant mean angular momentum in terms of the history effect along the Jaumann derivative.
We obtain Eq.~(\ref{eq:jaumannmodelintegrated}) via $R_{r\theta} = K b_{r\theta}^{(0)} + \lambda K b_{ij}^{(1)}$.

The time-local expression given by Eq.~(\ref{eq:jaumannexpandedmodel}) is also useful to understand the physical properties of the model.
Under Eq.~(\ref{eq:meanvelocitycondition}) in the inertial frame, we have
\begin{align}
    \frac{\mathscr{D} \widehat{S}_{i\ell}}{\mathscr{D} \widehat{t}} \widehat{W}_{\ell j} + \frac{\mathscr{D} \widehat{S}_{j\ell}}{\mathscr{D} \widehat{t}} \widehat{W}_{\ell i}
    = \frac{\mathscr{D}}{\mathscr{D} \widehat{t}} (\widehat{S}_{i\ell} \widehat{W}_{\ell j} + \widehat{S}_{j\ell} \widehat{W}_{\ell i}).
    \label{eq:simplifyswderivative}
\end{align}
If we assume that the similar condition to Eq.~(\ref{eq:simplifyswderivative}) holds even in Eq.~(\ref{eq:jaumannexpandedmodel}), the Reynolds stress up to $O(\lambda)$ and the first time derivative yields
\begin{align}
    R_{ij} = &
    \frac{2}{3} K \delta_{ij}
    - 2C_1 g \tau K S_{ij} 
    + 2C_1 (g\tau)^2 K \frac{\mathscr{D} S_{ij}}{\mathscr{D} t} 
    - 2C_1 C_{\mathscr{D}3} (g\tau)^2 K \left(S_{i\ell} W^\mathrm{A}_{\ell j} + S_{j\ell} W^\mathrm{A}_{\ell i} \right)
    \nonumber \\
    & + 4 C_1 C_{\mathscr{D}3} (g\tau)^3 K \frac{\mathscr{D}}{\mathscr{D} t}
    \left( S_{i\ell} W^\mathrm{A}_{\ell j} + S_{j\ell} W^\mathrm{A}_{\ell i} \right).
    \label{eq:jaumannmodelgeneral}
\end{align}
The shear stress component of Eq.~(\ref{eq:jaumannmodelgeneral}) yields Eq.~(\ref{eq:jaumannmodel}).

\bibliography{ref}

\end{document}